\documentclass{article}
\usepackage{emulateapj}
\usepackage{apjfonts}
\usepackage{psfig}
\def\Sec{${}^{\prime\prime}$\llap{.}}
\def\sec{${}^{\prime\prime}$~}

\def\Deg{${}^{\circ}$ }
\def\etal{{\it{et al. }}}
\def\fbss{$F_{BSS}$ }

\def\vmi{\hbox{\it V--I }}
\def\bmv{\hbox{\it B--V }}
\def\mag{$\,$mag }
\lefthead{JOHNSON ET AL.}
\righthead{OLD LMC CLUSTERS}

\begin{document}
\submitted{}
\title{Hubble Space Telescope Observations of the Oldest Star Clusters in the LMC}

\author{Jennifer A. Johnson{\altaffilmark{1}}, Michael Bolte}
\affil{UCO/Lick Observatory, University of California, Santa Cruz,
CA~95064; jennifer@ucolick.org, bolte@ucolick.org}
\author{Peter B. Stetson, James E. Hesser}
\affil{Dominion Astrophysical Observatory, Herzberg Institute of
Astrophysics, National Research Council of Canada, 5071 West Saanich Road,
Victoria, BC V8X 4M6, Canada; firstname.lastname@hia.nrc.ca}
\author{Rachel S. Somerville}
\affil{Racah Institute of Physics, The Hebrew University, Jerusalem, 91904, Israel; rachels@astro.huji.ac.il}
\altaffiltext{1}{Guest User, Canadian Astronomy Data Centre, which is
operated by the Herzberg Institute of Astrophysics, National Research
Council of Canada.}

\begin{abstract}
We present $V$, \vmi color-magnitude diagrams (CMDs) for three old star
clusters in the Large Magellanic Cloud (LMC):  NGC~1466, NGC~2257 and
Hodge~11.  Our data extend $\sim$ 3 magnitudes below the main-sequence
turnoff, allowing us to determine accurate relative ages and the blue
straggler frequencies.  Based on a differential comparison of the CMDs,
any age difference between the three LMC clusters is less than 1.5 Gyr.
Comparing their CMDs to those of M~92 and M~3, the LMC clusters,
unless their published metallicities are significantly in error, are the
same age as the old Galactic globulars.  The similar ages to Galactic
globulars are shown to be
consistent with hierarchial clustering models of galaxy formation.  The
blue straggler frequencies are also similar to those of Galactic
globular clusters.  We derive a true distance modulus to the LMC of
(m$-M)_0=18.46 \pm 0.09$ (assuming (m$-M)_0=14.61$ for M~92) using these
three LMC clusters.
\end{abstract}

\keywords{blue stragglers --- globular clusters: individual (NGC~1466, NGC 
2257, Hodge~11) --- Magellanic Clouds }

\section{Introduction}\label{intro}

Only in the Magellanic Clouds and the Milky Way Galaxy can we {\it
directly\/} measure the ages of old star clusters and thereby accurately
model the history of cluster formation. In the Galaxy, after decades of
research into dating techniques and collecting data for halo globular
clusters (including, very recently, globulars in the far reaches of the
halo using the Hubble Space Telescope and WFPC2 ({\it e.g.}, Harris \etal
1997; Stetson \etal 1999)), \markcite{har97}\markcite{stet98}we are
beginning to understand the details of the cluster age distribution. As
this story unfolds, a next step is to compare the early formation
history of the Galaxy with that inferred from the same measurement
techniques for star clusters in the Magellanic Clouds.

The most important questions to ask are (1)~whether the oldest Large
Magellanic Cloud (LMC) clusters are as old as the oldest Galactic
globular clusters (GGCs), and (2)~whether there is an age spread among
the most metal-poor LMC clusters.  The answer to the first question
tells us about the epoch of initial cluster formation for galaxies of
different mass and Hubble type.  The age spread among the old clusters
is one tracer of the early star formation history of the galaxy, and
indicates timescales important for collapsing or merging gas clouds.
Although the complete age distribution of the GGCs is not yet known,
the evidence to date suggests that the majority of the Milky Way's
halo clusters formed at the same time, with a small fraction of
clusters that formed up to $\sim$ 4 Gyrs later (Stetson, VandenBerg \&
Bolte 1996;\markcite{svb} although see Sarajedini, Chaboyer \&
Demarque 1997 for a different view).  What fraction of the co-eval
clusters formed in a single parent body or in smaller structures that
later merged in not clear.  The present results for the LMC also
suggest a mostly co-eval population of old clusters (Brocato
\etal 1996; Olsen \etal 1998)\markcite{bro96}\markcite{ol98}.

The first studies to attempt to measure ages for the old clusters in
the LMC appeared in the mid 1980s.  Studies of Hodge~11 (Stryker
et~al.\ 1984;\markcite{st84} Andersen, Blecha, \& Walker
1984),\markcite{abw} NGC~2257 (Stryker 1983;\markcite{st83} Hesser
et~al. 1984\markcite{hes84}), and several clusters in the series of
articles by Walker (referenced in Walker, 1993)\markcite{wh11} led the
way in studies of these sorts.  However the combination of large
distance to the LMC and problems with field-star contamination made
even relative age determinations with precision of a few Gyr or
better difficult.  Because of the superb resolution of the Hubble
Space Telescope (HST), color-magnitude diagrams (CMDs) can now be
measured in the cluster cores, which substantially reduces the impact
of field
star contamination.  With a modest effort using HST, stars three to
four magnitudes below the main-sequence turnoff can be measured with good
accuracy.

We have WFPC2 images in F555W and F814W of seven old LMC clusters
selected to have RR Lyrae stars or very blue horizontal-branch
morphology (see Table 1).  In this paper we present
color-magnitude diagrams for NGC~1466, NGC~2257 and Hodge~11, estimate
their ages relative to GGCs, and determine their blue straggler
specific frequencies.

\section{Observations and Data Reduction}
All of our clusters were observed in Cycle 5.\footnote[1]{Bolte, P.I.
Proposal Number 05897} Each cluster was imaged for 2 x 260s and 3 x
1000s in F555W ($\sim$ Johnson $V$) and 2 x 260s and 4 x 1000s in F814W
($\sim$ Cousins $I$).  The exposures were dithered to reduce the effect
of undersampling which is aggravated by subpixel quantum-efficiency
variations.  For NGC~1466, the cluster was centered on the PC.  Both
NGC~2257 and Hodge~11 were centered on WF3.  Figures 1a -- c
show the mosaicked images of our three clusters.
\begin{deluxetable}{llccc}
\tablenum{1}
\tablewidth{0pt}
\tablecaption{LMC Cluster Data}
\tablehead{
\colhead{Cluster}  & \colhead{[Fe/H]} & \colhead{ $M_V$} &
\colhead{Proj.~Radius} & \colhead {RR Lyrae?}
}
\startdata
Hodge 11 & $-2.06$ & $-6.99$ & 4.7$^\circ$ & no \nl
NGC 1466 & $-1.85$ & $-7.89$ & 8.4$^\circ$ & yes \nl
NGC 1841 & $-2.11$ & $-7.86$ & 8.4$^\circ$ & yes \nl
NGC 1786 & $-1.87$ & $-7.88$ & 2.5$^\circ$ & ? \nl
NGC 2210 & $-1.97$ & $-8.09$ & 4.4$^\circ$ & yes \nl
NGC 2257 & $-1.80$ & $-6.91$ & 8.4$^\circ$ & yes \nl
Reticulum& $-1.71$ & $-5.96$ & 11.4$^\circ$ & yes \nl
\enddata
\tablecomments{Data from Suntzeff {\it et al.} (1992)}
\end{deluxetable}
\begin{figure*}
\plottwo{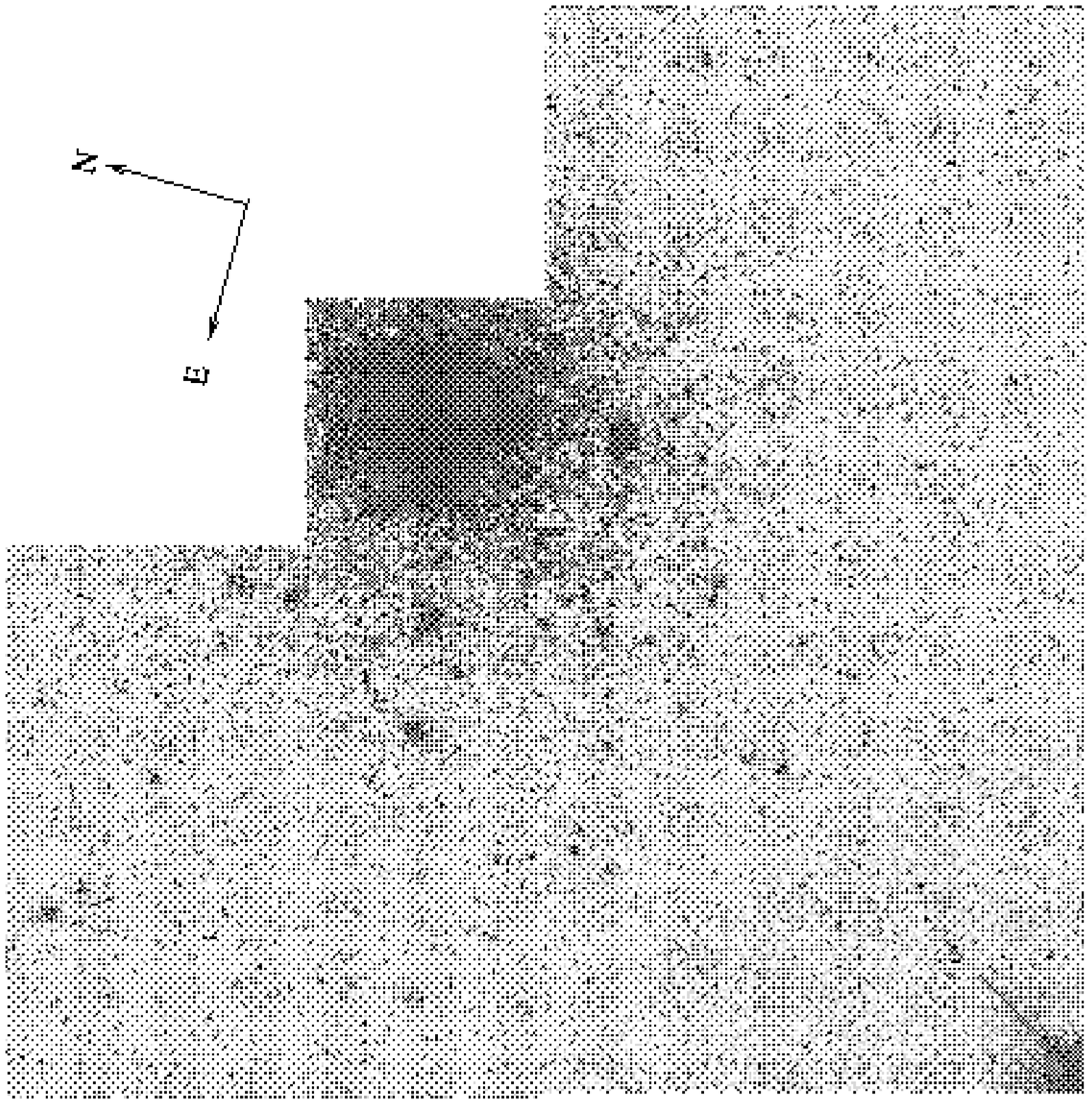}{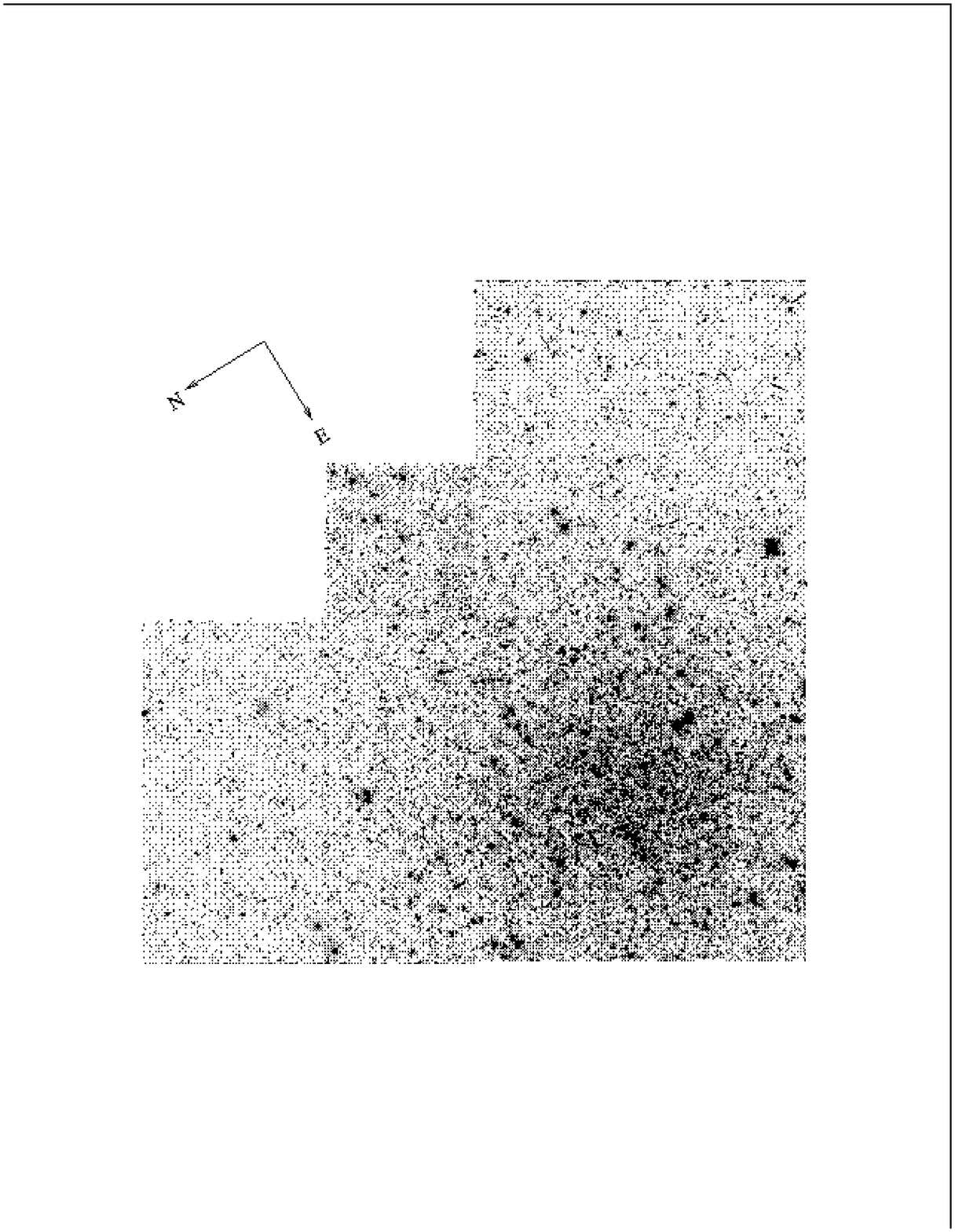}

\plottwo{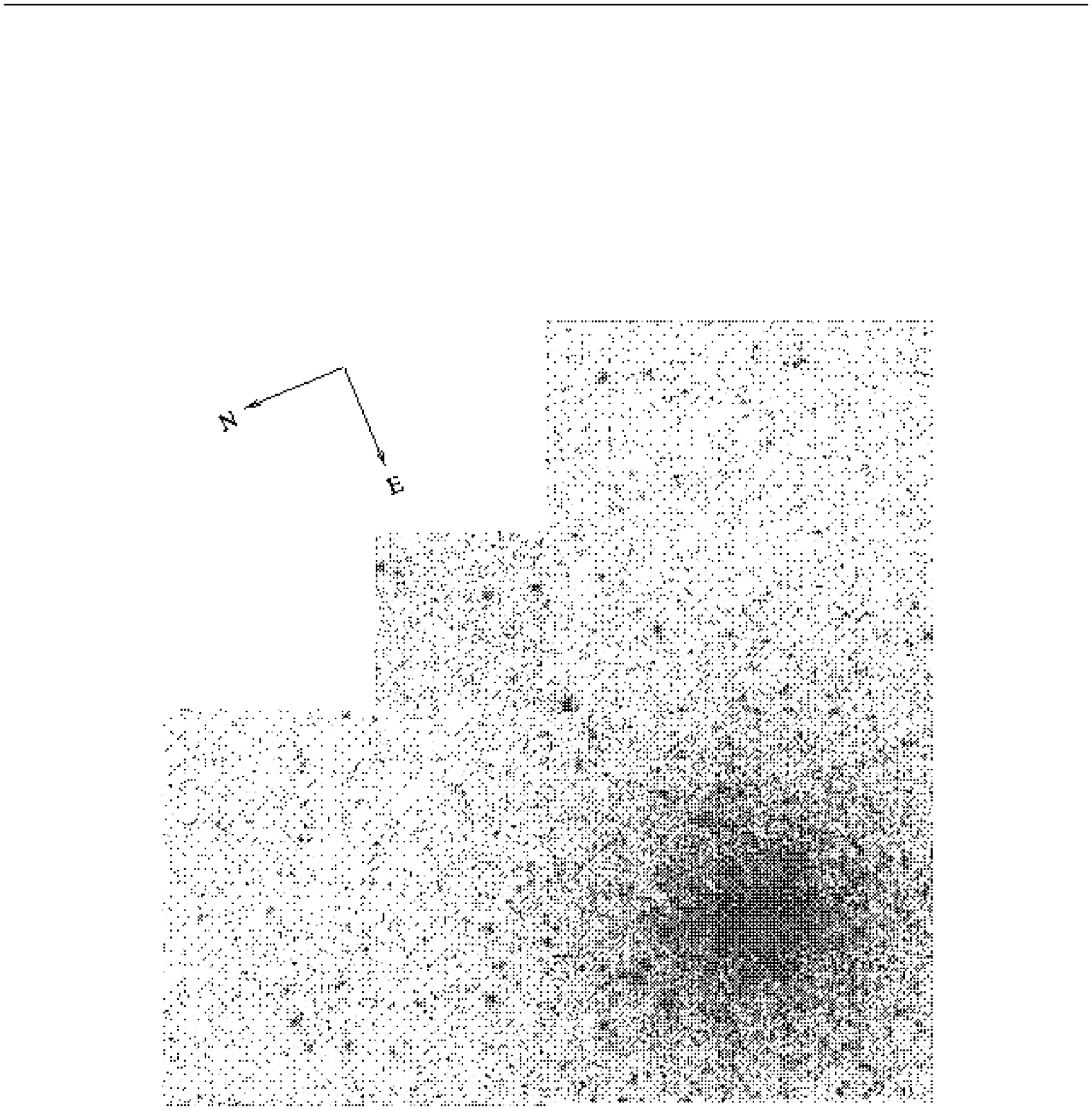}{f1d.ps}
\caption{Mosaicked $V$-band images of the NGC~1466 WFPC2 field (top-left),
the NGC~2257 WFPC2 field (top-right) and the Hodge~11 WFPC2 field (bottom-left)}
\end{figure*}

For our final reduction, we obtained recalibrated images from the
Canadian Astronomy Data Centre that had have been re-reduced using the most
appropriate biases, darks, flats and bad-pixel masks.  These
calibration frames were obtained closer in time to the object frames
than the ones originally used in the standard HST pipeline.  We then
masked the vignetted portions of the chips and corrected the pixel
values for changes in effective pixel area over the chips.

For the photometry, we used the ALLSTAR/ALLFRAME suite of programs
which do point-spread-function (PSF) fitting to determine magnitudes
(Stetson 1987, 1992,
1994)\markcite{stet87}\markcite{stet92}\markcite{stet94}.  The PSF
varies spatially across each chip and temporally as the HST focus
drifts.  We created a PSF by combining Tiny Tim (Krist
1995)\markcite{k95} PSFs and PSFs generated by DAOPHOT from our
images.  A DAOPHOT PSF for the inner three pixels of the model PSF was
created for each chip using the images that had the
best distribution of stars on that chip.  Because we lacked isolated, bright
stars to determine the wings of the PSF, we used the Tiny
Tim PSF beyond 3 pixels.  Judging by the widths of the horizontal and red
giant branches and by the lack of trends of aperture correction with
position, our PSFs did a reasonable, although not perfect, job of
mapping the spatial variations in the PSF.  We will return to this
when discussing our fiducial lines.

We set the weighting parameters ``profile error'' and ``percent
error'' to zero in ALLSTAR and ALLFRAME, as suggested by Cool \& King
(1995)\markcite{ck95}.  This eliminates the deviation of the pixels
from the PSF and errors in the flatfielding from the weighting scheme.
ALLSTAR and ALLFRAME essentially did intensity-weighted aperture photometry of
the inner 1.75-2 pixels, then added a correction for the amount of light
in the wings appropriate for the PSF at the star's position.

In detail, for each cluster we ran FIND and ALLSTAR on each frame
separately.  Next, we matched stars between frames using DAOMASTER
(Stetson 1993)\markcite{stet93} only including stars on the master
list which appeared on at least three frames.  This successfully
removed cosmic rays from the list.  This master list was input into
ALLFRAME which uses one star list for all frames, but determines
magnitudes for each frame individually.  Thus we have five measured
$V$ magnitudes and six $I$ magnitudes for most of the objects in each
cluster.

\begin{figure*}[ht]
\centerline{
\psfig{figure=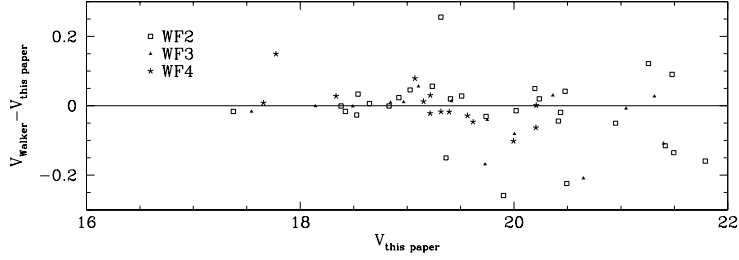,width=4.0truein,angle=-90}
}
\caption{Comparison of Walker's and our photometry for NGC~1466.  A
line is drawn at $\Delta mag=0$ for reference.}  
\end{figure*}

The ALLFRAME magnitudes required several corrections before we could
calibrate the data using the transformation equations of Holtzman
\etal (1995)\markcite{holtz95}.  We first needed to correct for the
difference between our PSF magnitudes and apertures of 0\Sec5, which
is the aperture size used by Holtzman \etal in deriving their
transformation equations.  To do this, we found cosmic ray hits on our images
by comparing each individual image with a cosmic-ray-cleaned image
created by taking the median of a registered stack of frames. Pixels
in each frame which deviated by more than 2$\sigma$ from the value in
the medianed image were flagged and the IRAF task FIXPIX was used to
interpolate over them.  We next selected a set of fairly isolated
stars on each frame and subtracted all other stars.  
Stars that had more than two flagged pixels within 0\Sec5 or any
pixels flagged within three pixels ($\sim$ two PSF fitting radii) of
their centers were rejected.  For each frame, the mean difference
between the PSF and 0\Sec5 aperture magnitudes for the culled list of
stars was determined.  This value was the aperture correction and was
added to all the PSF magnitudes for that frame.  

For the short exposures on the PC chip for NGC~2257 and Hodge 11, the
difference between aperture and PSF-based magnitudes was magnitude-dependent.  This is due to the WFPC2 charge-transfer efficiency (CTE) problems, since 
there were not
enough photons in the wings of the faint stars to fill up the charge
traps.  For these two clusters the PC has only a small fraction
of the total cluster stars, so we did not use it.  For NGC~1466, the PC
had a much higher background and the aperture corrections were constant
with magnitude.  We also note that we used sky annuli of 2\Sec0 to
2\Sec5 , instead of the 4\sec and 6\sec that Holtzman \etal used.  This
reduced the effects of badly subtracted neighbors in crowded regions
and uneven sky brightness across the cluster face.  However, there is a
difference between using the closer sky apertures and the more distant
ones due to contributions from the wings of the stars.  Based on our
model PSFs, this difference is only 0.001 mags -- much smaller than our
errors in the aperture corrections (see below).
\begin{figure*}[hb]
\centerline{
\psfig{figure=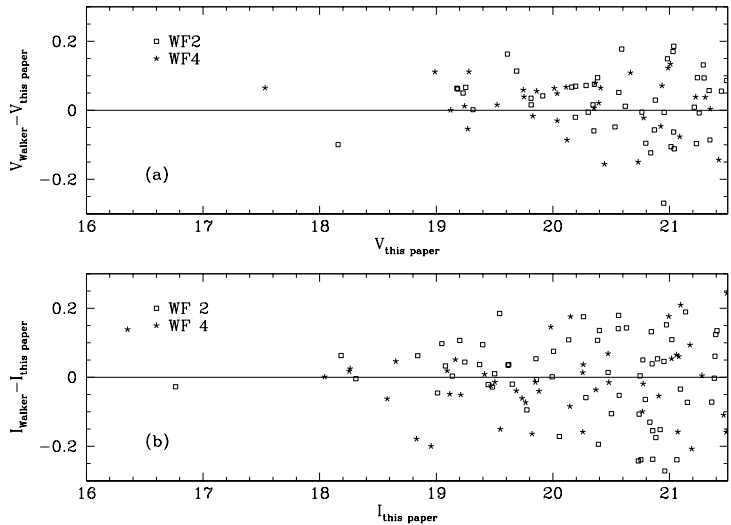,width=4.0truein,angle=-90}
}
\caption{Comparison of Walker's and our photometry for
Hodge~11.  (a) V magnitudes (b) I magnitudes.  A line is drawn at
$\Delta mag=0$ for reference.}
\end{figure*}

\begin{figure*}[hbt]
\centerline{
\psfig{figure=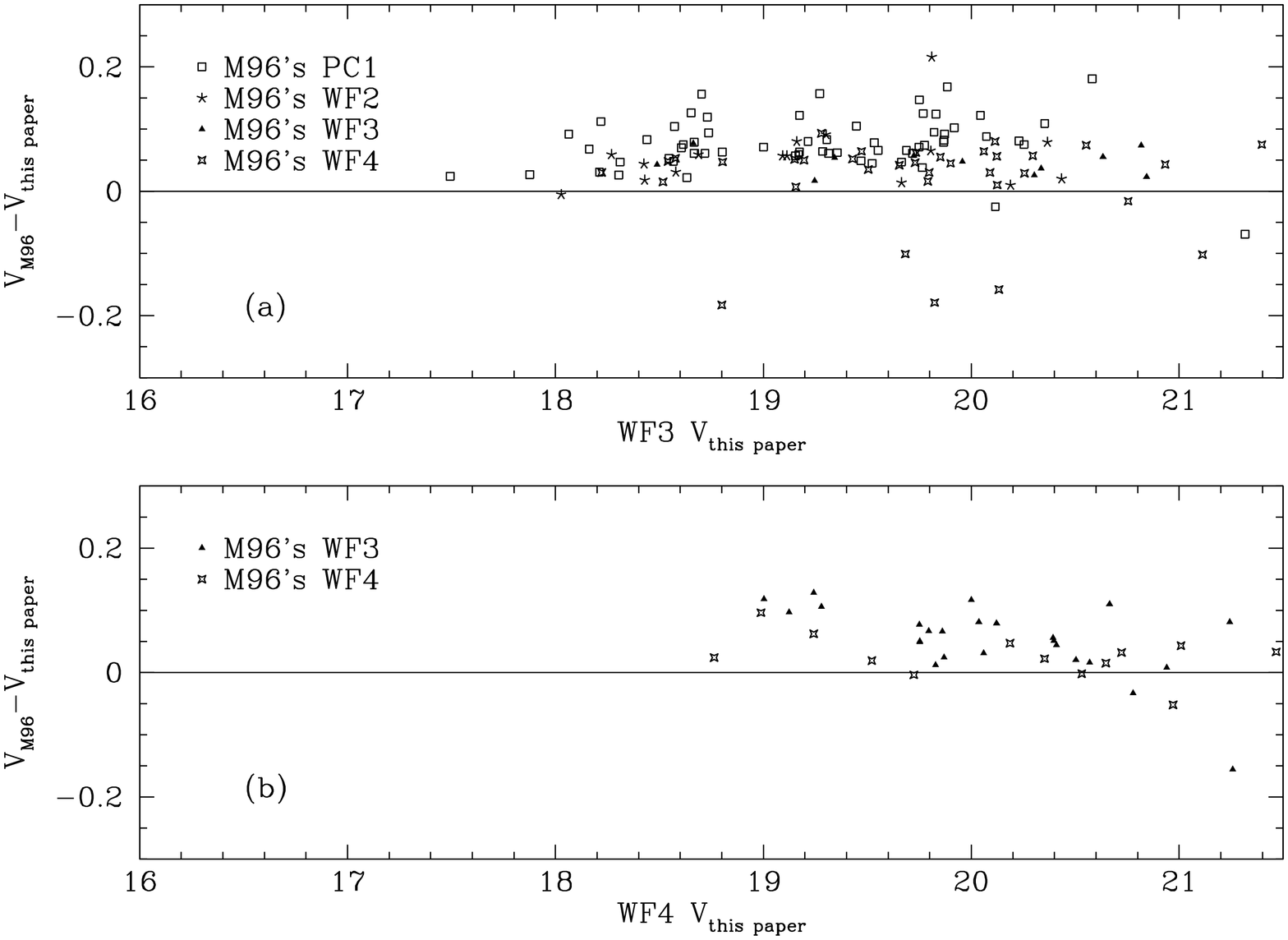,width=4.0truein,angle=-90}
}
\caption{Comparison of Mighell {\it et al.}'s (1996) and our photometry for
Hodge~11.  We plot the M96 stars found on our (a) WF3 and (b) WF4.  A
line is drawn at $\Delta mag=0$ for reference.} 
\end{figure*}
Within 0\Sec5, most of stars were affected by residuals from subtracted
neighbors and cosmic rays.  The RMS scatter in the aperture corrections
(aperture -- PSF magnitudes) for an individual frame ranged from $\sim$
0.02 to $\sim$ 0.04$\,$mag.  To see what random errors this observational
scatter introduced into our zeropoints, we used DAOMASTER to calculate
the average magnitude offset for each frame among all the stars common
to all frames in each filter + field combination.  After accounting for
the expected offset due to different exposure times, any remaining
offset is a result of errors in our aperture correction.  We are
assuming here that the PSF of WFPC2 did not change in the amount of
time it took to take images of one cluster ($\sim$ 4 hours).
We fit a Gaussian to the
distribution of all the offsets and found our 2$\sigma$ errors to be
0.014 mags.  Since DAOMASTER removes these offsets when averaging
magnitudes, essentially bringing all images to the zeropoint of one
frame, this is the error for our averaged magnitude zeropoints as
well.  We note that this error does not include any systematic error
that may result if 0\Sec5 apertures do not include a PSF-independent
fraction of the light.

After adding the aperture corrections, we corrected for the
CTE problem in the y-direction, using the
prescription of Whitmore \& Heyer (1997)\markcite{whit97} for 0\Sec5
apertures.  Our PSFs are not good enough to allow accurate
interpolation of saturated stars, so these stars were eliminated.  We
averaged our five F555W measurements and six F814W measurements using
DAOMASTER and kept only the stars with errors $<$ 0.08\mag in both
filters.  Finally, we calibrated the data using the Holtzman \etal
equations.  We list the photometry and positions for all our stars in Tables 
2a-c.  We used the first $V$ exposure to determine the
listed $X$ and $Y$ coordinates. (Tables 2a-c available from the first author)

In addition to the random errors caused by scatter in the aperture
corrections (0.014 $\,$mag), there is added uncertainty from the Holtzman
\etal zero points ($\sim$ 0.03$\,$mag), and from the Whitmore \& Heyer
corrections ($\sim$ 0.02$\,$mag) (see Stetson 1998\markcite{stet98}).
Adding these errors in quadrature
leads to an overall error of 0.04, most of it systematic, in our
zeropoints.  The systematic error in \vmi is more likely to to be
$\sim 0.02$, because the uncertainties in the Whitmore \& Heyer
corrections and probably in the Holtzman zero points are correlated
between $V$ and $I$.

\section{Comparison with Previous Photometry}

\begin{figure*}[htb]
\centerline{
\psfig{figure=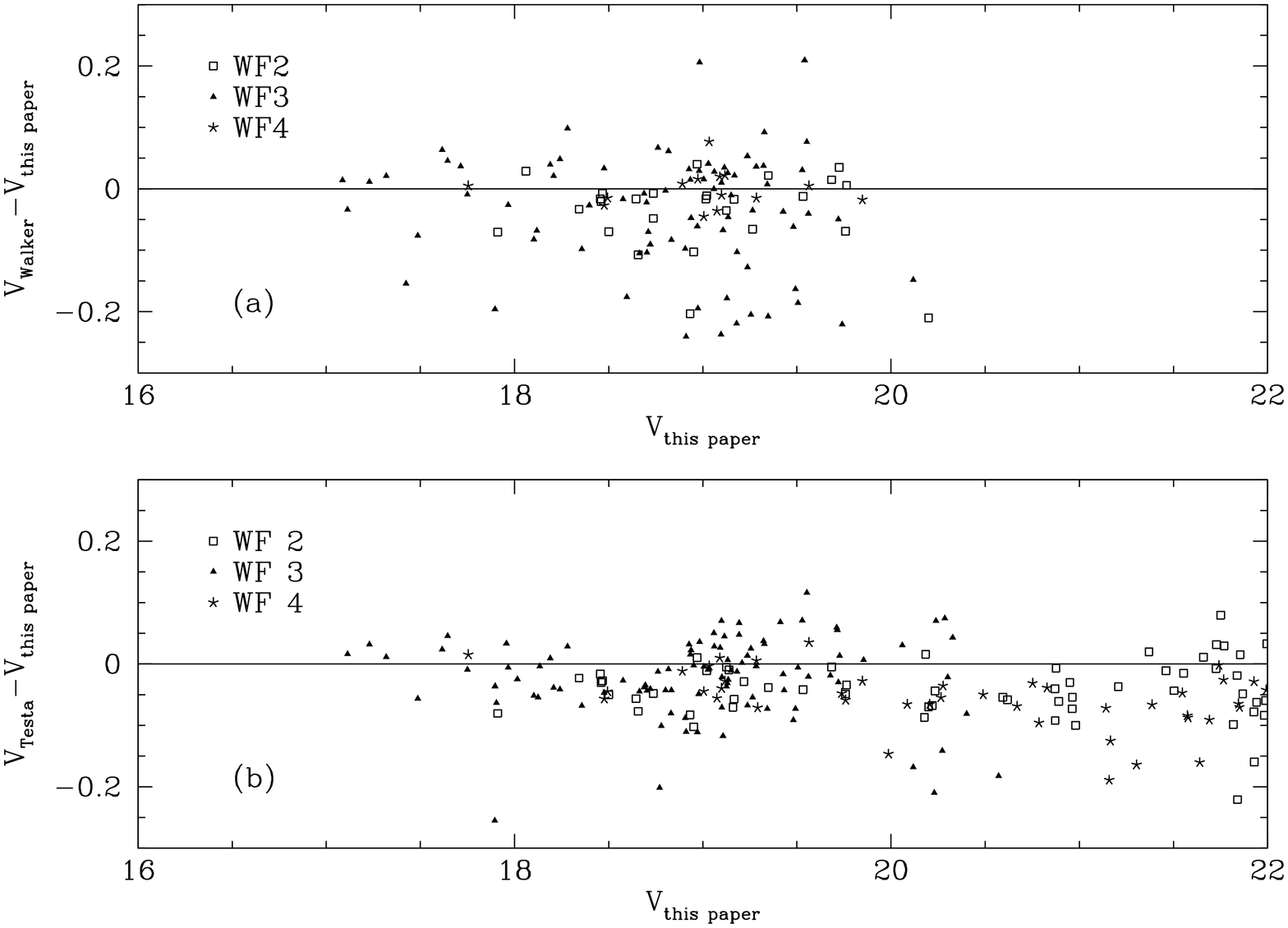,width=4.0truein,angle=-90}
}
\caption{Comparison of our photometry and (a) Walker's photometry (b) Testa 
{\it et al.}'s photometry for NGC~2257.  A line is drawn at $\Delta mag=0$ for 
reference.} 
\end{figure*}

Each of these clusters has at least one CCD-based photometric study
that we can compare with ours.  Our general
procedure for making comparisons was to find stars in common between
our study and the previous ones and eliminate all stars from the joint
sample where we detected a companion within 0\Sec5 in our HST images.
We also exclude stars within 1 magnitude of the detection limit of the
ground-based sample.  In each case our calculated offsets and errors
are tabulated in Table 3.

\begin{deluxetable}{lllclc}
\tablenum{3}
\tablewidth{0pt}
\tablecaption{Photometry Comparisons}
\tablehead{
\colhead{Cluster}  & \colhead{Author} &\colhead{Chip} & \colhead{Filter} &\colhead{Mag$_{p
revious}-$Mag$_{us}$} & \colhead{\# Stars}
}
\startdata
NGC 1466 & Walker 1992b & WF2 & V & $-0.007 \pm 0.018$ & \phn 32 \nl
NGC 1466 & Walker 1992b & WF3 & V & \phs $0.009 \pm 0.020$ & \phn 17 \nl
NGC 1466 & Walker 1992b & WF4 & V & $-0.008 \pm 0.016$ & \phn 16 \nl
Hodge 11 & Walker 1993  & WF2 & V & \phs $0.016 \pm 0.011$ & \phn 80 \nl
Hodge 11 & Walker 1993  & WF4 & V & \phs $0.039 \pm 0.014$ & \phn 53 \nl  
Hodge 11 & Walker 1993  & WF2 & I & $-0.007 \pm 0.026$ & \phn 87 \nl  
Hodge 11 & Walker 1993  & WF4 & I & $-0.007 \pm 0.021$ & \phn 54 \nl  
Hodge 11 & Mighell {\it et al.} 1996 & WF3 & V & \phs $0.051 \pm 0.005$ & 144 \nl
Hodge 11 & Mighell {\it et al.} 1996 & WF4 & V & \phs $0.032 \pm 0.007$ & \phn 46 \nl
NGC 2257 & Walker 1989 & WF2 & V & $-0.017 \pm 0.008$ & \phn 26 \nl
NGC 2257 & Walker 1989 & WF3 & V & $-0.027 \pm 0.013$ & \phn 83 \nl
NGC 2257 & Walker 1989 & WF4 & V & $-0.002 \pm 0.008$ & \phn 16 \nl
NGC 2257 & Testa {\it et al.} 1995 & WF2 & V & $-0.047 \pm 0.008$ & 102 \nl
NGC 2257 & Testa {\it et al.} 1995 & WF3 & V & $-0.012 \pm 0.005$ & 100 \nl
NGC 2257 & Testa {\it et al.} 1995 & WF4 & V & $-0.046 \pm 0.008$ & 109 \nl
\enddata
\end{deluxetable}

For NGC~1466, Walker (1992b)\markcite{w1466} published $B$, $V$ values
for stars with $V < $ 23.2\mag and further than 13\Sec6 from the
cluster center.  Figure 2 shows the differences between
photometries for the three WF chips.  (The PC had very few stars with
measurements by Walker.)  
The stars in common between the two studies are 
mostly at the faint end of Walker's study where he did
not quote errors.  So instead of finding a weighted mean, we found the
median $\Delta$mag.  Our estimate of $\sigma$ comes from half of the 
difference between the value $>$ 16\% of the $\Delta$mags and the
value $>$ 84\% of the $\Delta$mags.  This estimate of $\sigma$
is then reduced by square root of the number of measurements, to estimate
the standard error of the mean.

For Hodge~11, Walker (1993)\markcite{wh11} imaged in $V$ and $I$.
Because of the crowding from the ground, Walker excluded all stars
closer to the cluster center than 40$^{\prime\prime}$, so no stars from WF3 were
included in the comparison.  The $V$ and $I$ magnitude comparisons
(Figures 3a-b) show a large scatter, but no zeropoint
offset or trends with magnitude.
Mighell \etal (1996)\markcite{mig96} published $B$, $V$ data on
Hodge~11 obtained as part of a snapshot survey of Magellanic Cloud
clusters.  In
Figures 4a-b, we plot the difference between our WF3
and WF4 data and matching stars in the Mighell \etal photometry.  We
calculated a weighted average of difference, using $3\sigma$ clipping.

NGC~2257 is the sparsest cluster of the three and has
ground-based photometry in the cluster center.  Both Walker
(1989)\markcite{w2257} and Testa \etal (1995)\markcite{t95} have $B$,
$V$ data.  Testa \etal used Walker's photometry to calibrate
their data, but the Testa \etal
data extend two magnitudes fainter for stars further than 40\sec from
the cluster center.  Qualitatively, Figures 5a-b show that our
photometry is in reasonable agreement with previous efforts.  
Our zeropoint error estimate of 0.04\mag seems reasonable in light of
the comparison with previous photometry.  We note that relative age
determinations and blue straggler statistics are unaffected by
absolute scale concerns, although the reddenings and distances
determined will vary for different calibrations.  We see no
signs of non-linearity or color transformation problems that could
affect the results discussed in this paper.


\section{Color-Magnitude Diagrams} 

In Figures 6-8, we present the calibrated CMDs for the clusters.
\begin{figure*}[hbt]
\plotone{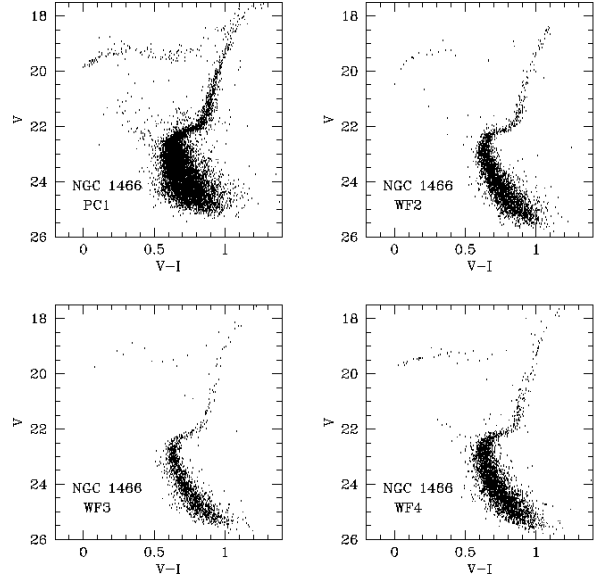}
\caption{NGC~1466 color-magnitude diagrams for the individual chips of
WFPC2.  The cluster is centered on the PC1.} 
\end{figure*}
\begin{figure*}
\plotone{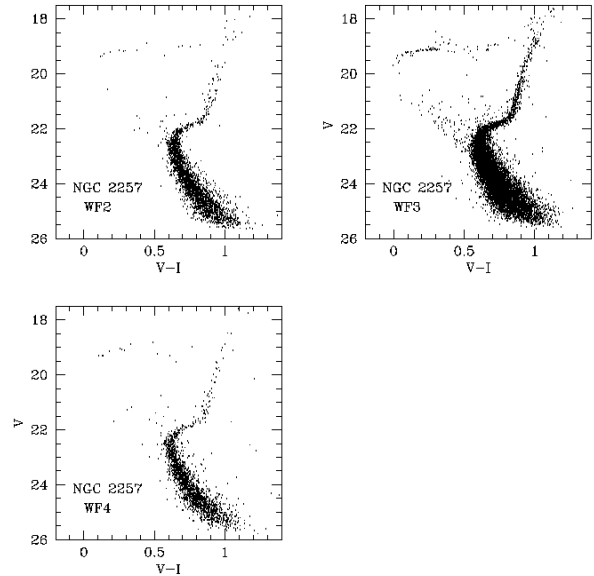}
\caption{NGC~2257 color-magnitude diagrams for the
individual chips of WFPC2.  The cluster is centered on the WF3.}
\end{figure*}
\begin{figure*}
\plotone{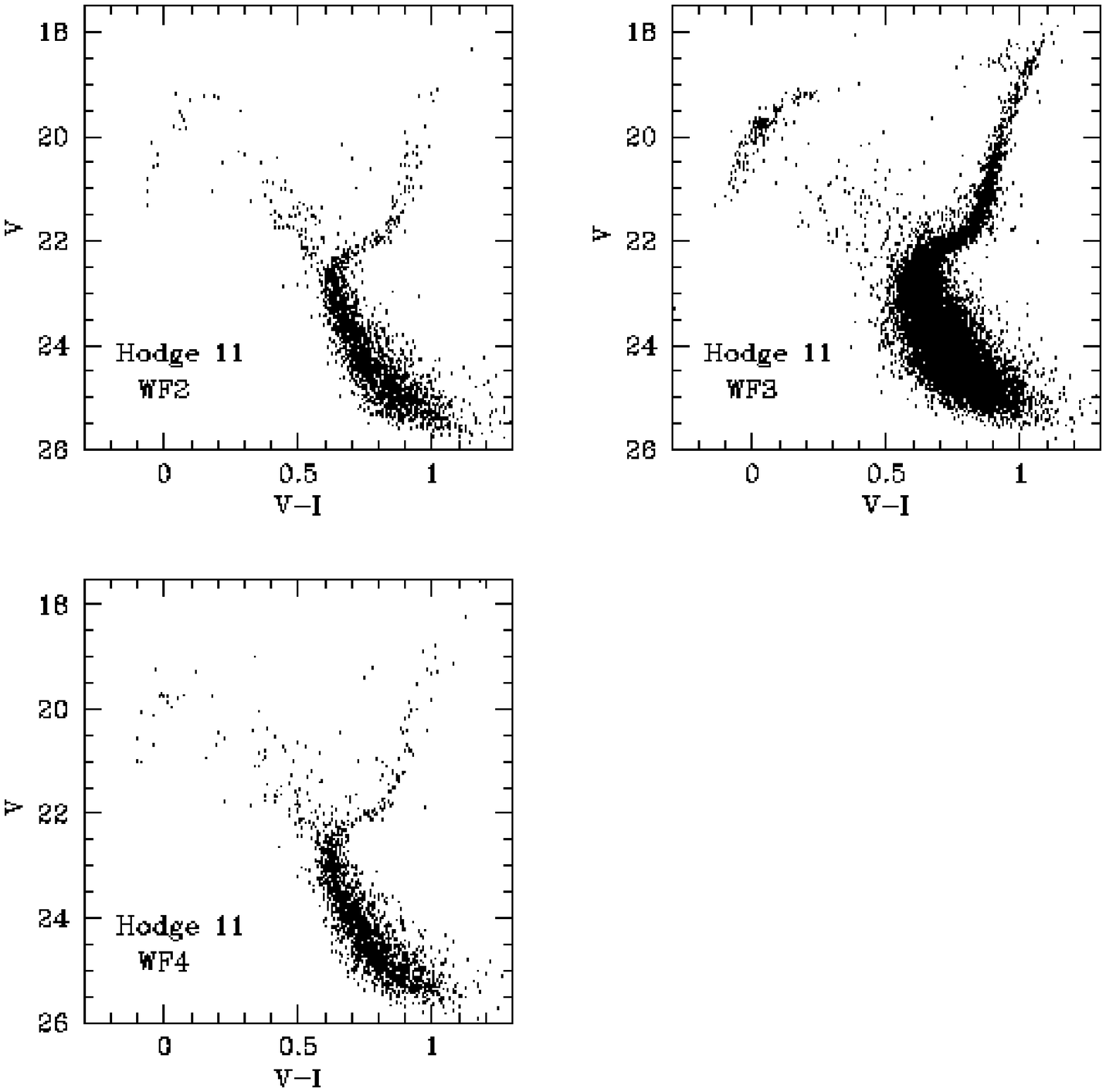}
\caption{Hodge~11 color-magnitude diagrams
for the individual chips of WFPC2.  The cluster is centered on the
WF3.}
\end{figure*}
Zero-point differences
cause slight offsets between the CMDs derived from the various chips
within a given cluster.  The largest shifts are $\sim$ 0.025\mag in color,
which can almost all be ascribed to the 0.015\mag random errors in the
aperture corrections in $V$ and $I$, with some error due to the Holtzman
gain ratios and Whitmore \& Heyer (1997) CTE corrections.
In all cases we have well-defined cluster
sequences from near the tip of the RGB down to $\sim$ 3\mag
fainter than the main-sequence turnoff (MSTO).  We determined the
fiducials for the stars on each chip using an algorithm described in
Sandquist \etal (1996)\markcite{sand96}.  The main-sequence fiducial
was determined by binning the data in magnitude and finding the mode
in color.  The red-giant-branch (RGB) fiducial was found by using the
mean color of magnitude bins.  The horizontal branch (HB) and subgiant
branch (SGB) could not be reliably fit by such methods, the HB because
of the large color dispersion of the RRLyr and the SGB because it was
neither vertical or horizontal.  In these two regions we found points
by eye, but the spread in magnitude in these areas is small, and this
procedure should not add large errors.  For chips with a smaller
number of stars, the bins used for finding the mode and mean were
larger, and the fiducials are noisier.  In Figures 9a-c, we plot the
fiducials from each chip for each cluster.  For NGC~2257 and Hodge~11,
the clusters which have their centers on WF3, we used a sample with
r~$>$~20\sec from the cluster center to determine the WF3 fiducial.
For NGC~1466, we used stars with r $>$ 9\Sec2 from the cluster center
for the PC1 fiducial.

\begin{figure*}[b]
\plottwo{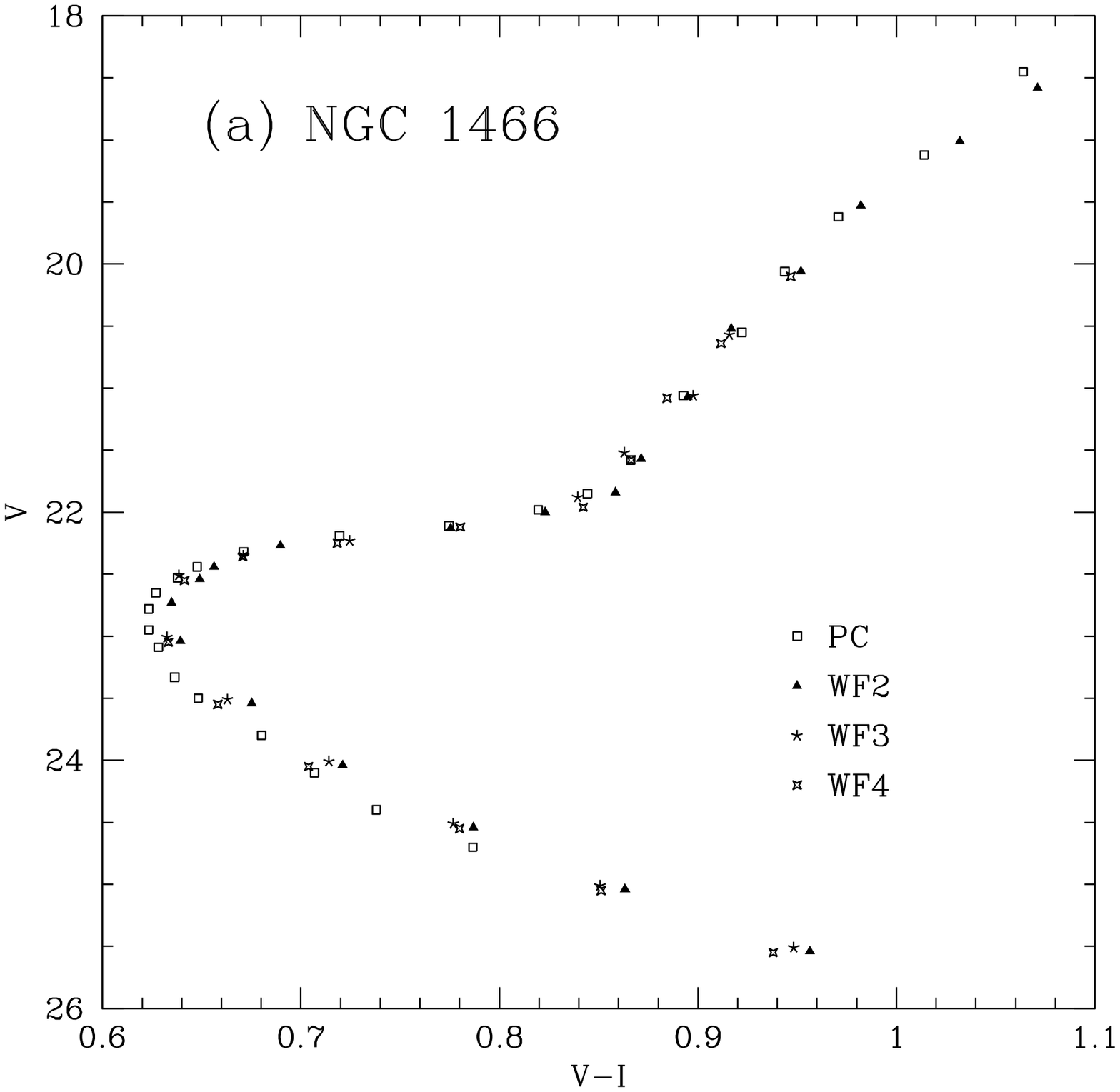}{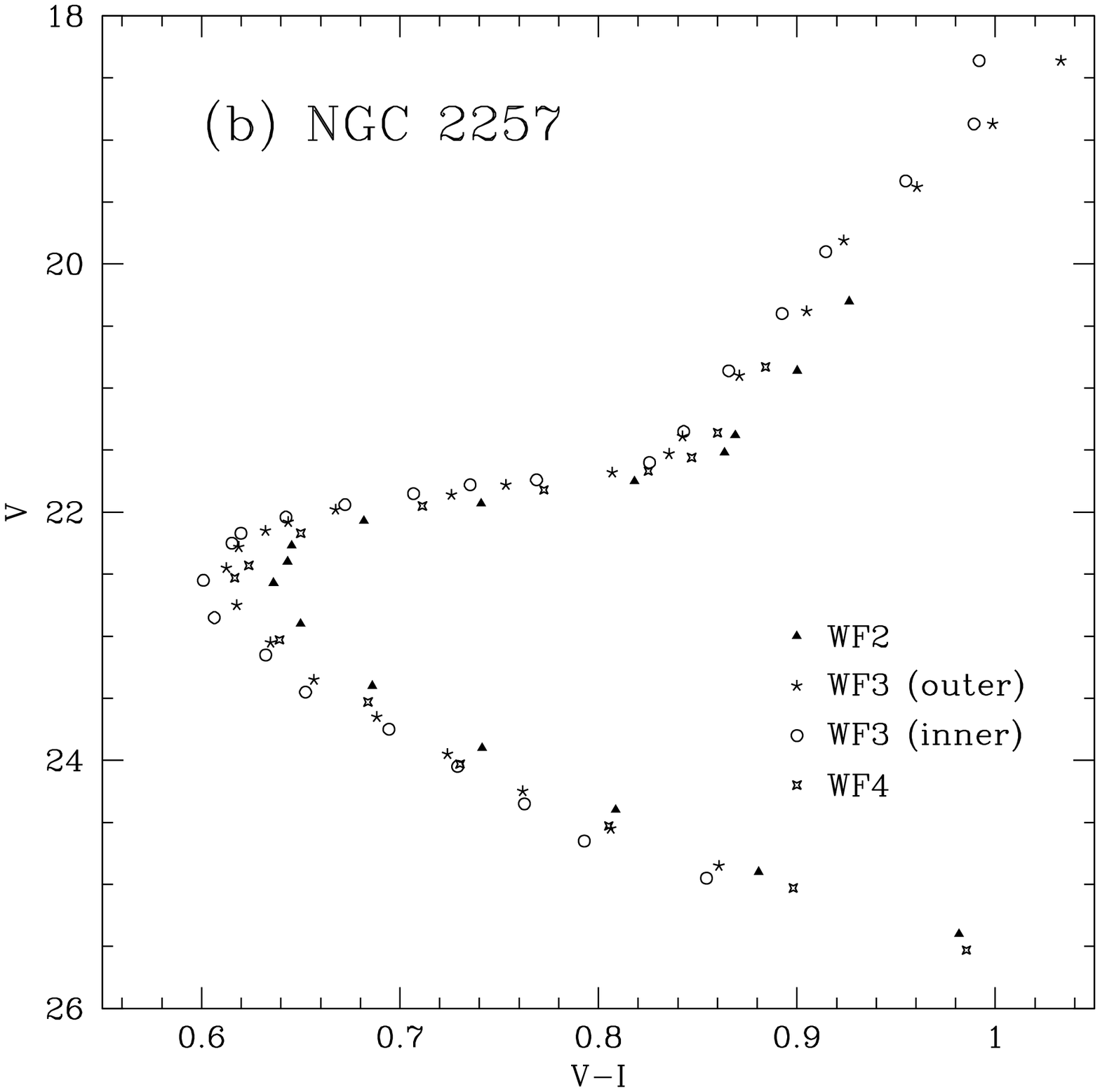}

\plottwo{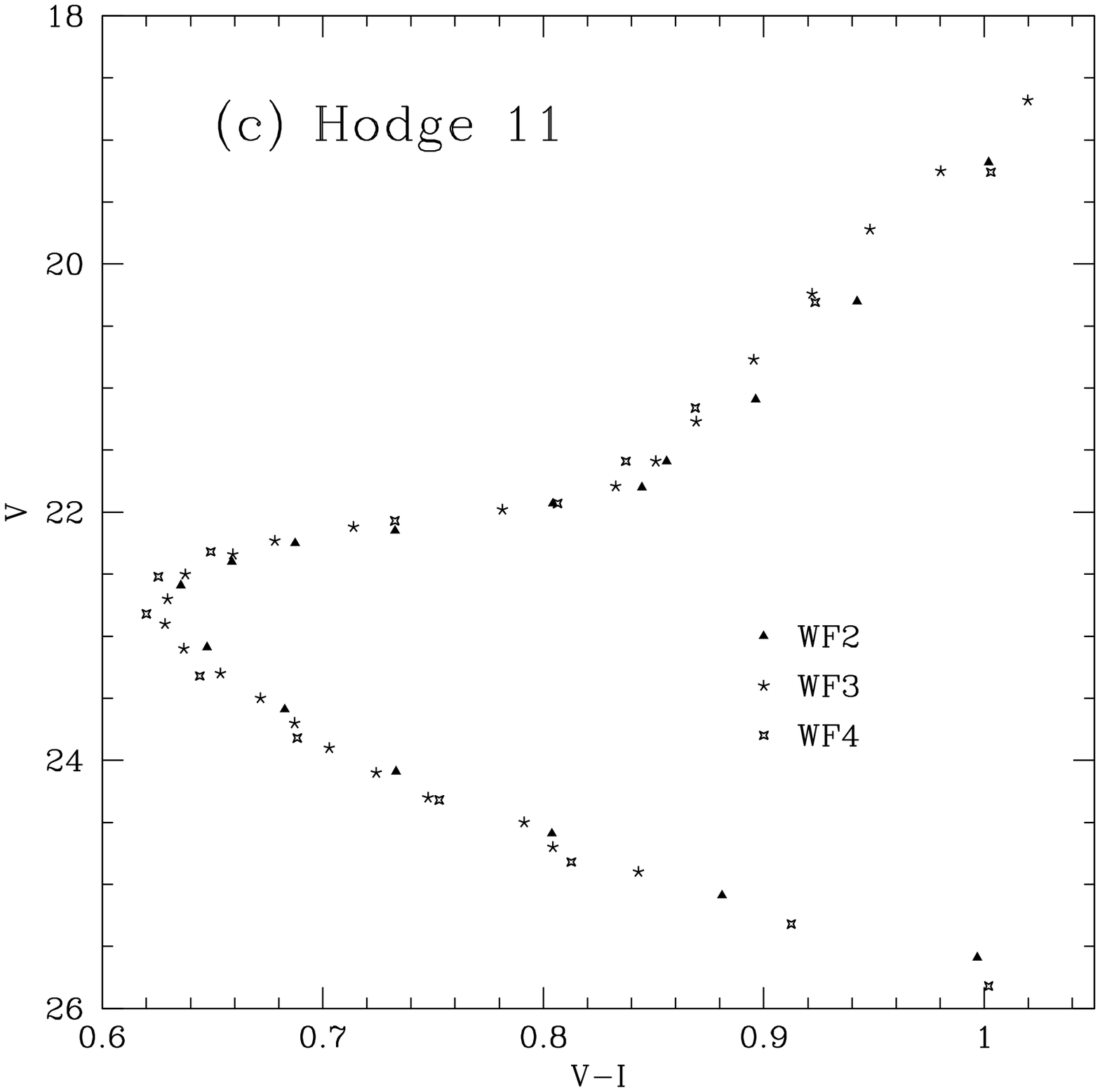}{f1d.ps}

\caption{The fiducials determined
separately from the data for each chip.  (a) NGC~1466 (b) NGC~2257 (c) Hodge
11.}  
\end{figure*}
\begin{deluxetable}{rr|rr|rr}
\tablenum{4}
\tablewidth{0pt}
\tablecaption{Cluster Fiducial Points}
\tablehead{
\multicolumn{2}{c}{NGC 1466} & \multicolumn{2}{c}{NGC 2257} &
\multicolumn{2}{c}{Hodge 11} \\
\colhead{$V$} & \colhead{$V-I$} & \colhead{$V$} & \colhead{$V-I$} &
\colhead{$V$} & \colhead{$V-I$}
}
\startdata
\multicolumn{2}{c}{MS-SGB-RGB} & \multicolumn{2}{c}{MS-SGB-RGB} & \multicolumn{2}{c}{MS-SG
B-RGB} \nl
24.70 &  0.787 &  24.85 &  0.861 &  24.90 &  0.843 \nl
24.40 &  0.738 &  24.55 &  0.806 &  24.70 &  0.804 \nl
24.10 &  0.707 &  24.25 &  0.762 &  24.50 &  0.791 \nl
23.80 &  0.680 &  23.95 &  0.724 &  24.30 &  0.748 \nl
23.50 &  0.648 &  23.65 &  0.688 &  24.10 &  0.724 \nl
23.33 &  0.637 &  23.35 &  0.657 &  23.90 &  0.703 \nl
23.09 &  0.628 &  23.05 &  0.635 &  23.70 &  0.687 \nl
22.95 &  0.623 &  22.75 &  0.618 &  23.50 &  0.672 \nl
22.78 &  0.623 &  22.45 &  0.613 &  23.30 &  0.654 \nl
22.65 &  0.627 &  22.28 &  0.619 &  23.10 &  0.637 \nl
22.53 &  0.638 &  22.15 &  0.632 &  22.90 &  0.628 \nl
22.44 &  0.648 &  22.08 &  0.644 &  22.70 &  0.630 \nl
22.32 &  0.671 &  21.98 &  0.668 &  22.50 &  0.638 \nl
22.19 &  0.720 &  21.86 &  0.726 &  22.34 &  0.659 \nl
22.11 &  0.775 &  21.78 &  0.753 &  22.23 &  0.678 \nl
21.98 &  0.820 &  21.68 &  0.807 &  22.12 &  0.714 \nl
21.85 &  0.844 &  21.53 &  0.836 &  21.98 &  0.781 \nl
21.58 &  0.866 &  21.39 &  0.842 &  21.79 &  0.833 \nl
21.06 &  0.893 &  20.90 &  0.871 &  21.59 &  0.851 \nl
20.55 &  0.922 &  20.38 &  0.905 &  21.27 &  0.869 \nl
20.06 &  0.944 &  19.81 &  0.924 &  20.77 &  0.896 \nl
19.62 &  0.971 &  19.38 &  0.961 &  20.24 &  0.922 \nl
19.12 &  1.014 &  18.87 &  0.999 &  19.72 &  0.948 \nl
18.45 &  1.064 &  18.36 &  1.033 &  19.25 &  0.980 \nl
\multicolumn{2}{c}{HB}  &  \multicolumn{2}{c}{HB} &  18.68 &  1.020 \nl
19.81 &  0.027 &  19.56 &  0.021 &  \multicolumn{2}{c}{HB} \nl
19.61 &  0.070 &  19.32 &  0.060 &  19.18 &  0.179 \nl
19.47 &  0.127 &  19.21 &  0.116 &  19.47 &  0.093 \nl
19.37 &  0.184 &  19.13 &  0.199 &  19.79 &  0.038 \nl
19.27 &  0.661 &  19.10 &  0.281 &  20.25 & $-0.010$ \nl
19.20 &  0.771 &  18.95 &  0.723 &  20.76 & $-0.050$ \nl
      &        &        &        &  20.80 & $-0.054$ \nl
\enddata
\end{deluxetable}

In our plot for NGC~2257, we include the fiducial
determined from the inner 20\sec  on WF3.  It is offset by a constant
color from the fiducial determined from the stars with r $>$ 20\sec .
We believe this is a manifestation of our less-than-perfect PSFs, which
do not fully account for the changes in the PSF over the face of the
chip.  However, examination of Figures 9a-c shows that this is not a major
concern, since the fiducials from each chip and each region look very
similar.  

For our final fiducials, we adopt the fiducial of the chip containing
the cluster center (Table 4).  We
also calculated the shifts needed to compensate for chip-to-chip
zero-point offsets
by matching the SGBs and RGBs of the individual chip fiducials.  These
shifts were applied to the data to create a master CMD of all stars on
all chips to use when finding blue stragglers.

\section{Relative Ages} \subsection{Metallicities} In order to
inter-compare the CMDs of the LMC clusters and to compare them with
those for the GGCs, we need to know their metallicities.  We will refer
all measurements to the Zinn \& West (1984; ZW84)\markcite{zw84}
scale.  The accuracy of this scale has been questioned by Carretta \&
Gratton (1997)\markcite{cg97}.  They determined [Fe/H] for individual
stars in 24 GGCs using a homogeneous analysis of equivalent widths of
Fe lines from high-resolution spectra.  Their correlation with the ZW84
metallicities is significantly non-linear, especially for intermediate
metallicities around $-1.5$.  Rutledge \etal (1997)\markcite{rut97}
compared metallicities based on $W'$, the reduced equivalent width of
Ca II triplet lines, with ZW84 and Carretta \& Gratton.  This estimate
of the metallicity also was related non-linearly to the ZW84 [Fe/H]'s,
but linearly to the high-resolution results.  This result is somewhat
ambiguous because of the dependence of $W'$ on log g and [Ca/Fe], but
shows again that the relative, in addition to the absolute, scale of
GGC metallicities has not been settled.  However, we are interested in
a small enough range that the ZW84 scale can be considered linear.
Also, for relative age studies, sensitivity to [Fe/H] uncertainties
decreases with decreasing Z and our results are not sensitive to [Fe/H]
errors $<0.3$ dex.

There are only a few studies of the abundances for  
the old LMC clusters. Cowley \& Hartwick (1982)\markcite{ch82}
used low-resolution spectra to measure the G-band strength, Ca II
break and the average strength of groups of weak Fe lines. They
calibrated indices against several GGCs and used the Zinn (1980a)
scale, which is almost identical to the ZW84 scale.  Olszewski \etal
(1991)\markcite{ols91} measured the strength of the Ca IR Triplet
lines for a large number of LMC clusters, including NGC~1466 and Hodge~11.  
They derive [Fe/H] via comparison with the Ca strength in stars 
in the GGCs NGC~288, NGC~1851, M~79 and NGC~7099 whose [Fe/H] were from ZW84.

For three stars in NGC~1466, Cowley \& Hartwick found [Fe/H]=$-2.0 \pm
0.2$.  Olszewski \etal (1991)\markcite{ols91} derived abundances for
two stars, LE4 and LE3, in NGC~1466 and found that LE4 had a
substantially lower [Fe/H] ($-2.48$) than LE3 ($-1.85$).  Walker
(1992b)\markcite{w1466} argued that since the apparent magnitude of
LE4 placed it significantly above the cluster sequences, it probably
had a close, bluer companion which weakened the \ion{Ca}{2} equivalent
widths.  Checking our HST images, we find LE3 is reasonably well
isolated, while LE4 is indeed blended with at least three other objects
that are closer than 0\Sec5.  Walker also estimated [Fe/H] via several methods
involving the properties of NGC~1466 RR-Lyrae stars and the
de-reddened colors of various [Fe/H]-sensitive points in the cluster
CMD. He found a consistent value of $-1.82 \pm 0.04$ from all of his
methods if E(\bmv) in the direction of NGC~1466 is $\sim 0.08$$\,$mag.

Cowley \& Hartwick measured three stars in Hodge~11
and derived a mean value of [Fe/H]$=-2.1 \pm 0.2$, consistent
with Olszewski {\it et al.}'s average from two stars of $-2.06 \pm 0.2$.
In this case, both stars had no companions within 0\Sec5 on our
HST frames.
Based on the de-reddened color of the giant branch in \vmi, Walker (1993),
using the calibration of Da Costa \& Armandroff (1990)\markcite{da90}, derives
[Fe/H]$=-2.0 \pm 0.2$ for E(\vmi)$=0.1$.

For NGC~2257, Cowley \& Hartwick measured a mean [Fe/H]$=-1.8 \pm 0.3$
from five stars.  Testa \etal measured the metallicity indicator
$(\bmv)_{0,g}$ (Sandage \& Smith 1966)\markcite{ss66} from their CMD
which leads to a value of $-1.86$ on the ZW84 scale.  Walker
(1989)\markcite{w2257} used the period-$A_B$ diagram of the RR~Lyraes
to estimate the metallicity as $-1.8 \pm 0.1$.  The agreement between
studies for each cluster is excellent.

We will adopt [Fe/H]=$-2.05$ for Hodge~11 and $-1.85$ for
NGC~1466 and NGC~2257. These measurements are on the
ZW84 scale, but in this [Fe/H] regime there is little
disagreement between ZW84 and any of the more recent scales.
The GGCs that we use for comparison are M~92 ($-2.25$) and M~3 ($-1.66$)
(ZW84).

It is of potential concern that the LMC clusters could have different
abundance ratios, such as [O/Fe], [C/Fe] and [$\alpha$/Fe], than their
Galactic counterparts.  However, Cowley \& Hartwick noted no
difference in the metallicities derived by comparing the strength of
weak Fe lines to those of Galactic clusters and the metallicities
found by looking at the \ion{Ca}{2} break and the G-band, which are
dominated by the abundance of the $\alpha$ elements. In the absence of
any evidence to the contrary, we will assume the same trends in
abundance ratios as are in GGCs.

\subsection{Ages:  CMD comparisons}

Figures 10-12 show the fiducials for the LMC clusters compared to $V$,
 \vmi fiducials for M~92 and M~3 (Johnson \& Bolte
1998)\markcite{me98}.  
\begin{figure*}[hbt]
\centerline{
\psfig{figure=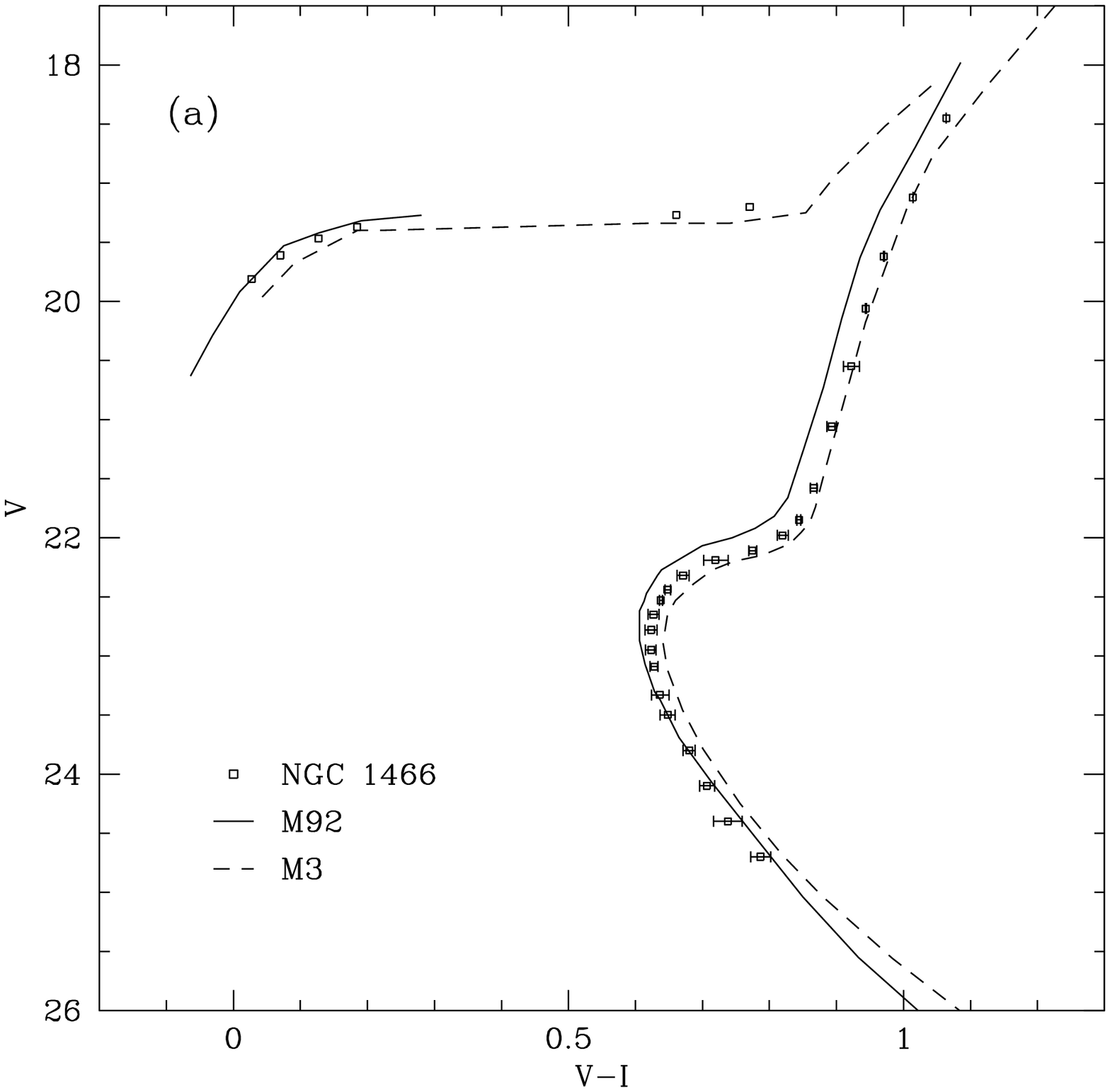,width=2.5truein,angle=0}
\psfig{figure=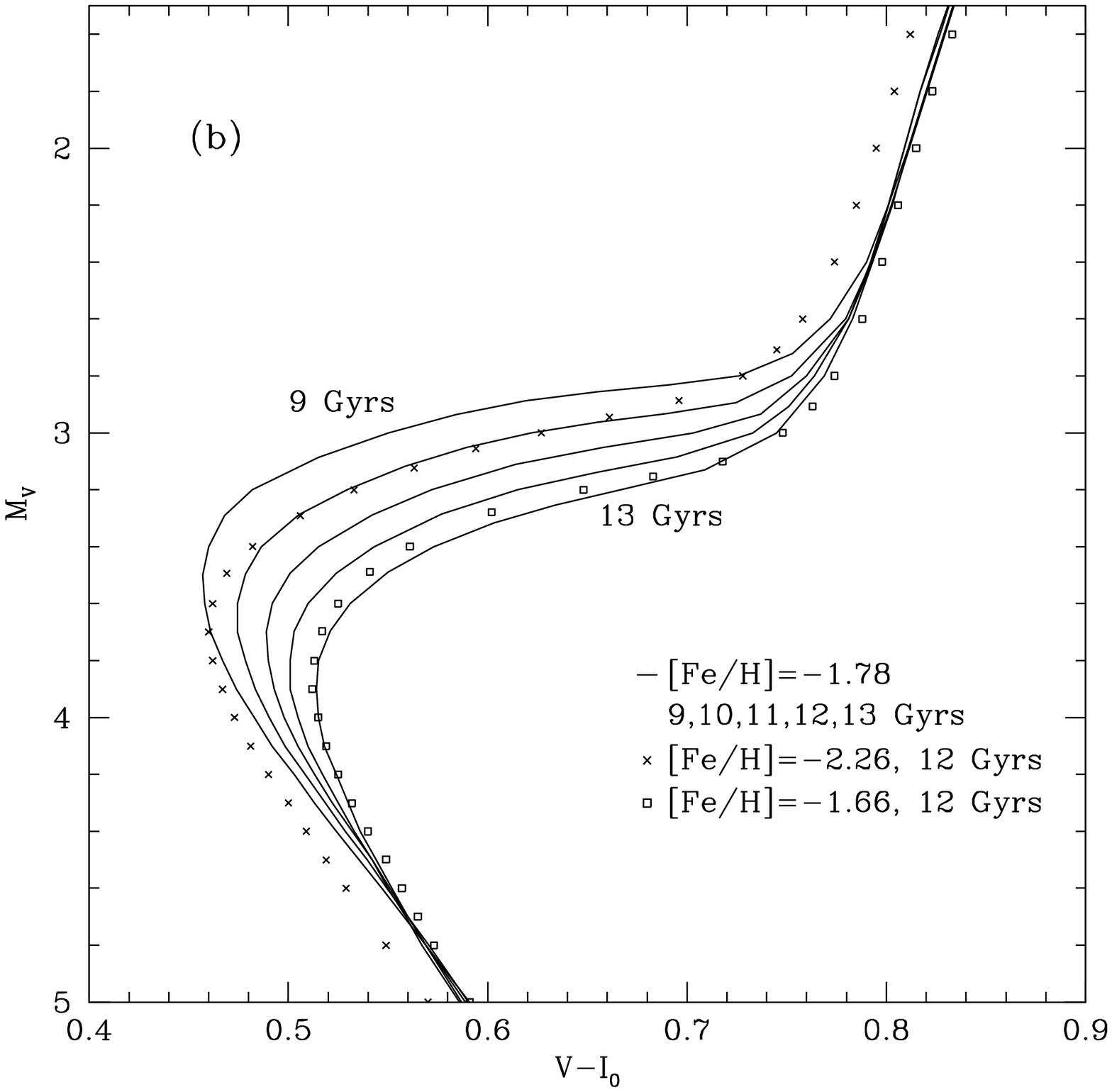,width=2.5truein,angle=0}
}
\caption{(a) NGC~1466's fiducial
compared with M~92 and M~3.  We have two estimates of the error in the
NGC~1466 points.  First, we determined the average deviation of the PC
fiducial from the WF fiducials in Figure 9a.  Second, we estimate the
random error in the color of each bin to be 0.005 magnitudes.  We adopted the
larger number as our error.  The $V$ errors in the HB magnitude
are smaller than the boxes used.  However, the M~92 and M~3 HBs have
larger, but unknown, errors because of the relatively few stars that
define them (see Johnson \& Bolte 1998 for more discussion).  (In particular,
the blue end of M3's HB is defined by one star.)  (b)
Bergbusch \& Vandenberg (1992) isochrones compared in a similar
manner.  Isochrones of 12 Gyrs are plotted for [Fe/H]=-2.26, -1.78 and
-1.66.  To illustrate the effect of age, [Fe/H]=-1.78 isochrones are
also plotted for 9, 10, 11, and 13 Gyrs.  ([Fe/H]=-1.78 is more
metal-rich than the values (c.f. Table 1) for our clusters, so the correct
isochrones would be bluer and brighter.)}  
\end{figure*} 

We accounted for the differences in distance and
small differences in abundance by
registering the HB magnitudes at the blue edge of the RR Lyr gap so
that $\Delta V_{HB} = -0.20 \Delta [Fe/H]$.  We accounted for the
differences in reddening by making the relative colors of the RGB
agree with Bergbusch \& VandenBerg (1992) theoretical isochrones of the correct metallicity.  In
Figure 10a, the NGC~1466 fiducial line fits between M~92 and M~3 from
the RGB (of course this is guaranteed by our procedure for correcting
for E(\vmi) on the upper RGB) down through the MS as expected for a
cluster with a metallicity intermediate between M~92 and M~3 and the
same age as these two clusters.  In Figure 10b, we show how well we
can determine an age difference from such plots.  We have taken
Bergbusch \& VandenBerg (1992)\markcite{berg92} (BV92) isochrones for
[Fe/H]=$-2.26$, $-1.78$ and $-1.66$. On this plot, we represent M~92 and M~3
by 12 Gyr isochrones of [Fe/H]=$-2.26$ and $-1.66$.  Then we
plot [Fe/H]=$-1.78$ isochrones for 13, 12, 11, 10 and 9 Gyrs.  The
differences in the turnoff region are striking and show that NGC~1466
has the same age as M~3 and M~92 to $\leq$ 1.5 Gyr.  For NGC~2257 and
Hodge~11, Figures 11-12 show the comparisons, and we reach similar
conclusions.
\begin{figure*}
\centerline{
\psfig{figure=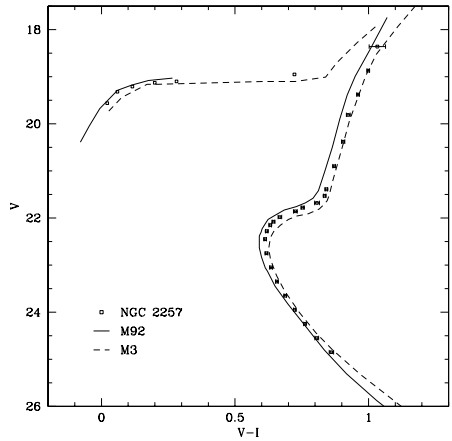,width=2.5truein,angle=0}
}
\caption{NGC~2257's fiducial compared with M~92 and M~3.  See the caption
for Figure 10 for a discussion of the errors.}
\end{figure*} 
\begin{figure*} 
\centerline{
\psfig{figure=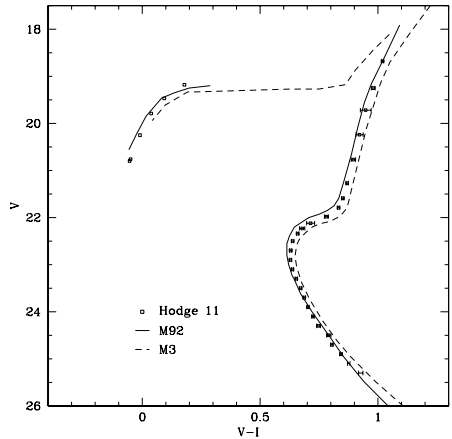,width=2.5truein,angle=0}
}
\caption{Hodge~11's fiducial compared with M~92 and M~3.  See the caption
for Figure 10 for a discussion of the errors.} 
\end{figure*}

\subsection{Ages:  Color-Difference Method}

We can quantify the age differences between the GGC and the LMC
clusters and among the LMC clusters themselves by measuring the color
difference between the MSTO and the base of the RGB (VandenBerg,
Bolte, \& Stetson 1990 (VBS)\markcite{vbs}; Sarajedini \& Demarque
1990\markcite{sd}).  In practice, we do this by shifting the cluster
fiducials horizontally until the colors of the MSTO match and
vertically until a magnitude reference point matches.  Since the MSTO
does not have a well-defined magnitude, VBS suggested using the point
on the MS that is 0.05 redder than the MSTO color ($V_{+0.05}$).  All
other things held constant, an older cluster will have a shorter SGB
and, after MSTO registration, its RGB will lie to the blue of that for
a younger cluster.  Unfortunately, the length of the SGB also changes
with metallicity.
  While this is a relatively small effect in \bmv (VBS), it is larger
when \vmi colors are used.  In \vmi, the [$\alpha$/H] is the crucial
abundance to know (VandenBerg, private communication) and in all cases
we have a good relative metallicities from Ca lines.  To determine the
magnitude of the metallicity effect requires the use of theoretical isochrones,
and in particular, theoretical colors.  Since colors are among the
most uncertain quantities predicted by theory, we will use two
separate sets of isochrones to evaluate our results and their
dependence on the choice of isochrones.

Our LMC fiducials are in many ways an ideal dataset to use with this
method.  They were all taken with the same instrument and have the
same data quality.  They have low metallicities within 0.2 dex of each
other, so the effects on the colors are minimized.  We measured the
MSTO color by fitting a parabola to the individual points in that
region.  We then interpolated between our fiducial points on the MS to
determine $V_{0.05}$.  In Figure 13 we have registered the LMC
cluster fiducials using these two points.  
\begin{figure*}
\plotone{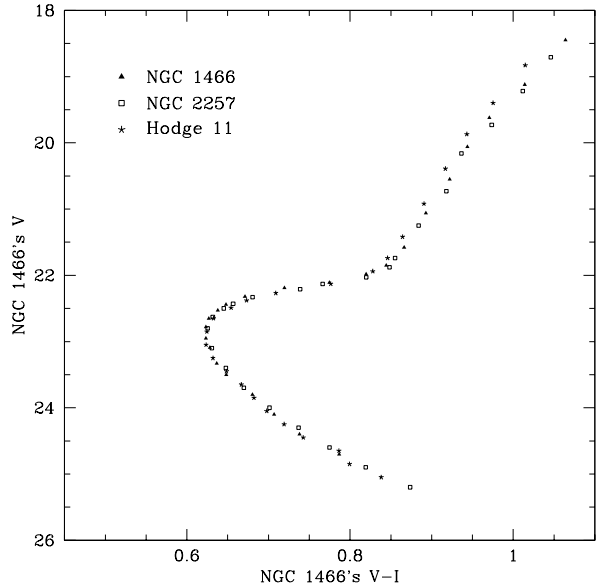}
\caption{NGC~1466, NGC~2257 and Hodge~11 fiducials compared using the 
color-difference method.  The remarkable agreement in the position of
their RGBs, especially for NGC~1466 and NGC~2257, argues against a large age
difference between the clusters.}
\end{figure*}
The sequences for NGC~1466
and NGC~2257 are indistinguishable.  Hodge~11's RGB lies to the blue
of the other two.  With the color-difference method in \vmi, the more
metal-poor Hodge~11 should be redder than NGC~1466 and NGC~2257, so
this indicates that Hodge 11 is somewhat older than the other two.  As
shown below, the derived age difference is not large.

Next, we quantified the age difference between the LMC clusters and
the GGCs based on any shifts of the postions of the registered RGBs.
We first fit the GGCs RGBs from the base to the magnitude of the HB
with a straight line.  Next, we calculated the weighted average of the
color offset between these lines and the individual RGB stars below
the HB of each of our LMC clusters. In Figure 14, we
show an example of the method where we have registered M~92 and
NGC~2257 using the two fiducial points.  
\begin{figure*}
\plotone{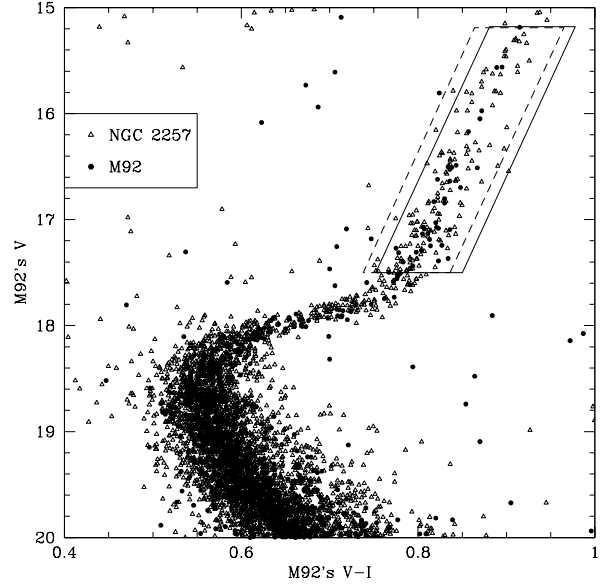}
\caption{The color-difference method applied to NGC~2257 and M~92.  We
plotted the individual stars for each cluster.  The solid trapezoid
shows the M 92 stars used to calculate the straight line fit to the RGB.  The dashed trape
zoid encloses the NGC~2257 stars used to 
calculate the RGB shift relative to the M~92 line.}  
\end{figure*}
Based on VBS, the {\it
mean} offset for a cluster gives the age difference (compared to M~92)
and the standard error of the mean offset gives an estimate of our
$\delta$ age precision.  The mean offsets for each cluster are
recorded in Tables 5a-b as $\delta$(\vmi).
 
\begin{deluxetable}{llllll}
\tablenum{5a}
\tablewidth{0pt}
\tablecaption{$\delta(V-I)$ Quantities}
\tablehead{
\colhead{Cluster} & \colhead{$(V-I)_{TO}$} & \colhead{$V_{+0.05}$} & \colhead {$\delta(V-I
)$} 
& \colhead {$\Delta$ Age (Gyrs)} & \colhead {$\Delta$ Age (Gyrs)} \\
 \colhead{} & \colhead {} & \colhead{} & \colhead {} & \colhead{(BV92)}
& \colhead{(C95)} 
}

\startdata
M 92 & $0.553 \pm 0.008$ & $19.45 \pm 0.05$ & \phs $0.000 \pm 0.014$ & \phs $0.0 \pm 0.7$ 
 & \phs $0.0 \pm 0.7$ \nl
NGC 1466 & $0.623 \pm 0.005$ & $23.73 \pm 0.05$ & $-0.012 \pm 0.009$ & $-0.8 \pm 0.4$ & $-
1.0 \pm 0.5$ \nl
NGC 2257 & $0.610 \pm 0.005$ & $23.38 \pm 0.05$ & $-0.008 \pm 0.009$ & $-1.0 \pm 0.4$ & $-
1.2 \pm 0.5$ \nl
Hodge 11 & $0.628 \pm 0.005$ & $23.58 \pm 0.05$ & $-0.021 \pm 0.009$ & $+0.3 \pm 0.4$ & $+
0.3 \pm 0.5$ \nl
\enddata
\end{deluxetable}
\begin{deluxetable}{llllll}
\tablenum{5b}
\tablewidth{0pt}
\tablecaption{$\delta(V-I)$ Quantities}
\tablehead{
\colhead{Cluster} & \colhead{$(V-I)_{TO}$} & \colhead{$V_{+0.05}$} & \colhead {$\delta(V-I
)$} 
& \colhead {$\Delta$ Age (Gyrs)} & \colhead {$\Delta$ Age (Gyrs)} \\
 \colhead{} & \colhead {} & \colhead{} & \colhead {} & \colhead{(BV92)}
& \colhead{(C95)} 
}

\startdata
M3 & $0.595 \pm 0.005$ & $19.92 \pm 0.05$ & \phs $0.000 \pm 0.009$ & \phs $0.00 \pm 0.4$ &
 \phs $0.00 \pm 0.5$ \nl 
NGC 1466 & $0.623 \pm 0.005$ & $23.73 \pm 0.05$ & \phs $0.008 \pm 0.009$ & $+0.24 \pm 0.4$
  & $+0.36 \pm 0.5$ \nl
NGC 2257 & $0.610 \pm 0.005$ & $23.38 \pm 0.05$ & \phs $0.011 \pm 0.009$ & $+0.10 \pm 0.4$
 & $+0.21 \pm 0.5$ \nl
Hodge 11 & $0.628 \pm 0.005$ & $23.58 \pm 0.05$ & $-0.002 \pm 0.009$ & $+1.28 \pm 0.4$ & $
+1.56 \pm 0.5$ \nl
\enddata
\end{deluxetable}

While the standard error of the mean represents the error in our
measured shifts caused by scatter on the RGBs, it does not take into
account errors in our choice of \vmi$_{MSTO}$ and $V_{+0.05}$.  
The error in $V_{+0.05}$ caused by scatter
on the MS has only a small effect on the color shift of the RGB since the
RGB is almost vertical.  A shift of 0.05\mag in $V_{+0.05}$
results in a color shift of 0.0025$\,$mag. The more significant error is
in $\vmi_MSTO$.  This color could change by an amount of
order \vmi $\sim 0.003$\mag if we made plausible alterations in the fitting
range and the acceptable color error for our parabolic fit.  We
conservatively estimate our observational error in \vmi$_{MSTO}$ as
0.005$\,$mag.  Unfortunately, an error in the color also results in an error in
our choice of $V_{+0.05}$.  These two errors work in the same
direction.  For example, if our MSTO color is too red, our $V_{+0.05}$
will be too faint.  We will then shift our fiducial too far to the
blue to get our red MSTO color to match and too bright (and hence too
blue) to get our $V_{+0.05}$ to match.  Taking into consideration the
slopes of the MS and the RGB, we find that our 0.005\mag error in color
results in an additional 0.0035\mag shift in the RGB color due to the
correlation of these errors.  Our total random error is therefore
0.009$\,$mag.  Finally, we note that since we calculate the
difference for our three LMC clusters relative to one GGC at a time,
the errors in the GGC points will result in a systematic shift.
Comparing the results obtained for M~92 in Table 5a with those for M~3
in Table 5b will provide an estimate of the size of that effect.

To convert our measured $\delta$(\vmi) and errors into age differences
and uncertainties, we used two sets of isochrones: Bergbusch \&
VandenBerg (1992) (BV92)\markcite{berg92} and the new Yale Isochrones
(Chaboyer \etal 1995)\markcite{chab95} (C95).  We determined the predicted
shifts in RGB color as functions of metallicity and age after
registering the isochrones in the same manner as our data.  For $-1.66
>[{\rm Fe/H}]>-2.26$, we found that the color shift predicted by BV92
depends on the metallicities by the following relation:

\begin{equation}
\delta(\vmi)=-0.069(\Delta[Fe/H]).
\end{equation}
Next, comparing sets of isochrones with the same
metallicity, but with ages ranging from 9 to 13 Gyrs, we found that
the RGB shift is related to the age difference (in Gyr) by
\begin{equation}
\delta(\vmi)=-0.021(\Delta Age).
\end{equation}

We first removed the known shift due to metallicity differences from
our results in Tables 5a-b.  The remaining shift is due to age
differences.  Using our age-$\delta$(\vmi) relation above, we calculated
the age differences relative to the GGCs and included those values in
Tables 5a-b.  Our age-$\delta$(\vmi) relation is valid only for ages
between 9 and 13 Gyrs.  Our use of younger isochrones to calibrate the
RGB shifts is not crucial for our discussion of relative ages, but we
note that the choice of an older age range would increase the implied
age difference for a given $\delta$(\vmi).  However, the errors would
also increase, and we would be left with the same conclusion.

We performed the same procedure using the C95
isochrones with two exceptions.  First, the C95 isochrones are not
$\alpha$-enhanced, so we followed their recommendation and determined
the appropriate solar-scaled isochrones by using the formulation of
Salaris, Chieffi, \& Straniero (1993)\markcite{scs} and choosing the
relative $\alpha$ abundances to be $4\times$ solar.  Second, we found a
noticeable trend in the steepness of the slope of the
age-$\delta$\vmi relation depending on the metallicity of the
isochrones.  However, this affected our ages only at the level of 0.05
Gyrs, so we adopted the slope determined from the [Fe/H]$=-1.66$
isochrones.  Our equations for $\delta$(\vmi) versus metallicity and
age are

\begin{equation}
\delta(\vmi)=-0.080(\Delta[Fe/H]), \quad\quad {\rm and}
\end{equation}

\begin{equation}
\delta(\vmi)=-0.020(Age)
\end{equation}
The C95 isochrones have a steeper dependence on metallicity, but
the overall agreement between BV92 and C95 is quite good.  The
calibration of the color-difference method will need to updated again
as newer isochrones become available, but there is no evidence that
the theoretical calibration will be an overwhelming source of error.
Based on the $\delta$(\vmi) method, we conclude that the LMC clusters
are the same age as the GGCs to within 1.5 Gyrs.  

\subsection{Ages:  $V_{BTO}$}

The relative ages of GCs can also be calculated by using the
difference in $V$ magnitude between the HB and the MSTO (the $\Delta
V$ method) (Sandage 1982)\markcite{san82}.  Because luminosities can
be predicted more reliably than temperatures, this method has been
often put forward as the most robust of the relative age estimators.
However, it is relatively more uncertain for clusters
with very blue HB morphologies such as Hodge~11 for at least two
reasons.  First, it is difficult to assign an apparent brightness to
the HB and to correct for HB-star evolution in blue-HB clusters.
Second, as pointed out recently by Sweigart (1997)\markcite{sw97} and
Langer, Bolte \& Sandquist (1999)\markcite{lbs99}, there is reason to
suspect that some blue HB stars have had helium mixed into their
envelopes resulting in higher HB luminosity and, for clusters
containing He-mixed stars, an artificially inflated $\Delta V$ value.
In addition, our M 3 and M 92 data do not have many HB stars, which
limits the precision of our HB magnitudes.  Finally, determining the
magnitude of the MSTO is uncertain, since the color-magnitude diagram
is vertical at that point.

Chaboyer \etal (1996a)\markcite{chab96a} proposed a modification that
moved the lower reference point to the subgiant branch, 0.05\mag redder
than the MSTO color ($V_{BTO}$).  This reduces the errors, since the
CMD cluster sequence is sloped at that point and a fiducial point can
be determined with more confidence.  To find $V_{BTO}$, we used the
mean magnitude of the stars in a box on the subgiant branch centered on
our fiducial at $(V-I)_{MSTO}+0.05$.  The box was 0.04\mag wide in
color and 0.15\mag wide in $V$.  The standard error of the mean for
those points within the box provides one estimate of the error, but
probably underestimates it because there is additional confusion at the
blue end of the SGB due to the scatter of stars from the MS to the
red.  This makes the SGB wider in magnitude and makes $V_{BTO}$
dependent on our exact box size in vertical direction.  Our total
errors are therefore closer to $\sim 0.03$$\,$mag.  For the HB magnitude
of NGC~1466 and NGC~2257, we found the average and the standard error
of the mean for our HB stars at the blue end of the RRLyr gap.  This is
the method by which we found the HB magnitude for our GGC data, and
avoids the problem of determining $<M_V(RR)>$ with our limited time
resolution.  We used the HB of NGC~2257 as a template to extend Hodge
11's extremely blue horizontal branch to the red.  We ignored the two
reddest points of the Hodge~11 HB as likely being either evolved stars
or field interlopers.  The Hodge~11 HB magnitude derived in this manner
is very uncertain.  Table 6 summarizes our values for $V_{HB}$ and
$V_{BTO}$.  

\begin{deluxetable}{lcllll}
\tablenum{6}
\tablewidth{0pt}
\tablecaption{$V_{BTO}$ Quantities}
\tablehead{
\colhead{Cluster} & \colhead{$(V-I)_{TO}$} & \colhead{$V_{BTO}$} & \colhead{$V_{HB}$} & \c
olhead {$M_{V(BTO)}$} & \colhead {$\Delta$ Age (Gyrs)}
}

\startdata
M3       & 0.595 & $18.65 \pm 0.05$ & $15.75 \pm 0.05$ & $3.26 \pm 0.07$ & \phs $0.0 \pm 0
.8$ \nl
M92     &  0.553 & $18.14 \pm 0.05$ & $15.18 \pm 0.08$ & $3.22 \pm 0.09$ & $-0.4 \pm 0.8$ 
\nl
NGC 1466 & 0.623 & $22.33 \pm 0.03$ & $19.32 \pm 0.01$ & $3.34 \pm 0.03$ & \phs $0.8 \pm 0
.4$  \nl
NGC 2257 & 0.610 & $22.04 \pm 0.03$ & $19.10 \pm 0.01$ & $3.27 \pm 0.03$ & \phs $0.1 \pm 0
.4$ \nl
Hodge 11 & 0.628 & $22.25 \pm 0.03$ & $19.11 \pm 0.1$  & $3.44 \pm 0.10$ & \phs $2.0 \pm 1
.5$\nl
\enddata
\end{deluxetable}

To convert $V_{BTO}$ to $M_V(BTO)$, we adopt the
relationship from Gratton {\it et al.}'s (1997)\markcite{gr97}
analysis of the Hipparcos data,

\begin{equation} M_V(HB)=0.17([Fe/H]+1.5) + 0.39.  
\end{equation} 
In choosing this relationship for the magnitude of the HB, we are
implicitly choosing absolute ages for the globular clusters of about
12 Gyrs.  As we discuss in \S 5.3, this affects the precision of our
results, but it does not affect our conclusions.  

Chaboyer \etal
(1996a)\markcite{chab96a} provide the conversion between $M_{V(BTO)}$
and age for $V, I$ data for five metallicities.  The relevant
equations for our discussion are

\begin{equation}
t_9 = 70.7 - 48.5M_{V(BTO)} + 9.3M_{V(BTO)}^2  \quad\quad\quad\quad 
[Fe/H]=-2.5
\end{equation}

\begin{equation}
t_9 = 71.4 - 47.5M_{V(BTO)} + 8.8M_{V(BTO)}^2  \quad\quad\quad\quad 
[Fe/H]=-2.0
\end{equation}

\begin{equation}
t_9 = 86.3 - 53.5M_{V(BTO)} + 7.3M_{V(BTO)}^2. \quad\quad\quad\quad 
[Fe/H]=-1.5
\end{equation}

For all these clusters, we used the equation valid for [Fe/H]=$-2.0$
to derive ages (Table 6).
This is too metal-poor for NGC 1466, NGC~2257 and M~3 and too
metal-rich for M~92, and metallicity does affect the magnitude of the
MSTO.  To estimate the size of this effect, we used M 3's $M_{V(BTO)}$
to find its age using equations (7) and (8), which bracket M 3's
[Fe/H].  We then assumed a linear relationship between $\Delta$[Fe/H]
and $\Delta$Age and found that we should make M~3's age younger by
$\sim$ 0.6 Gyrs to account for the 0.20 dex difference in metallicity.
This is still consistent with being co-eval with NGC~1466 and
NGC~2257.  A similar calculation using equations (6) and (7) suggests
that M~92's age should be 0.8 Gyrs older.  The errors shown for
$\Delta$~Age in Table 6 include only the observational errors and not
possible errors from incorrect metallicities or distances.  Hodge 11
appears to be 2 Gyrs older by this method, but there is a large
additional component in the error due to the uncertainty in setting
the HB level.  These calculations confirm our result in the previous
sections.

Each method of calculating relative ages suffers from both systematic
and random uncertainties.  However, by using a combination of
these methods, we can hope to determine the allowable age differences.
Despite potential problems with finding HB magnitudes, relying on
theoretical isochrones for calibration of age differences in color and
magnitude, calculating fiducials for clusters with different crowding
conditions and using the $HST$ calibrations, we consistently find
no large age differences between M~92 and M~3 on one hand and NGC~1466,
NGC~2257 and Hodge~11 on the other.   Unless one of these methods is
found to suffer from a serious flaw, a reasonable statement is that
the GGC and these LMC clusters have the same age $\pm$ 1.5 Gyr.  
However, we find that Hodge 11 consistently looks a little older.
This may be due to an actual age difference, but the $\Delta V$ measurement
in particular could be due to the mixing of helium into the envelope, as
discussed above.  The extremely blue HB of Hodge 11, bluer than the
more metal-poor M 92, would agree with either an age difference or
helium-mixing.

\subsection{Distances}

The distance to the LMC is an important and controversial quantity
(see {\it e.g.}, Westerlund 1997 for a review). The LMC provides a
testing ground for consistency between the Cepheid and RR Lyrae
distance scales. Recent estimates for the LMC distance modulus have
ranged from (m$-M$)$_0=18.05$, based on observations of ``red clump''
stars (Stanek, Zaritsky \& Harris, 1998),\markcite{szh} to
(m$-M$)$_0=18.7$, based on Cepheid properties (Feast \& Catchpole
1997)\markcite{fc97}.  Fernley \etal (1998)\markcite{fern1998} used Hipparcos
proper motions for Galactic RR Lyrae stars and statistical parallax to
derive (m$-M$)$_0=18.26$. The most recent estimate based on the
illumination of the SN1987a ring is (m$-M$)$_0<18.44$ (Gould \& Uza
1998)\markcite{gu98}.  Generally, the RR-Lyrae-based distance to the
LMC has been smaller than that set by Cepheid observations (see {\it
e.g.}, Walker 1992a)\markcite{wrr}.  Adding to the confusion is the
possibility that RR-Lyrae stars in the LMC may have systematically
different luminosities than Galactic RR Lyraes (Walker 1992a; van den Bergh
1995)\markcite{vdb95}.

We can use our data to examine these issues by determining the
relative distance moduli between the LMC clusters and M~92 via (1)
matching HB levels and (2) matching the unevolved main-sequence (MS)
positions.  Our data extend far enough down the MS that we can use
method (2).  Because of the steep slope of the main sequence, the
color offset due to reddening and systematic errors in \vmi must be
removed before the MSs can be matched vertically.  We do this by
shifting the RGBs to their expected relative positions and then
adjusting the distance moduli to match the MSs.  This is an iterative
process, and not as precise as knowing the reddening separately.  We
find that our range of acceptable fits leads to errors in the distance
moduli of $\sim$ 0.05$\,$mag.  As can be seen in Figures 10 through
12, {\it we would have derived identical relative distances if we had
matched the HBs of the clusters}. This latter observation strongly
suggests that there is no large difference ($\sim$ 0.3\mag) in the
luminosity of HB stars between the LMC clusters and M~92.  We then
corrected these distance moduli for the differential extinction
between M~92 and these clusters.  We adopted the reddenings of Walker
(1992a) for our LMC clusters and $E(\bmv)=0.02$\mag for M~92 (Harris
1996).  These true distance moduli between M~92 and the LMC clusters
are listed in Table 7 as $\mu_{M92,GC}$.

\begin{deluxetable}{llccccll}
\tablenum{7}
\tablewidth{0pt}
\tablecaption{Distance Moduli}
\tablehead{
\colhead{Cluster} & \colhead{$A_V$} & \colhead{$\mu_{M92,GC}$} & 
\colhead{$\mu_{\odot,GC}$} & \colhead{$\mu_{\odot,LMC}$} &
\colhead {$\mu_{\odot,GC}$} & \colhead{$V_{HB}$} & \colhead{``$V_{HB}$''}\\
\colhead{} & \colhead{} & \colhead{} &\colhead{Pont}
 &\colhead{Pont} &\colhead{Gratton} & \colhead{} & \colhead{Walker}
}
\startdata
M92        & 0.06 & .... & 14.61 & ....  & 14.74 & 15.18 & .... \nl
NGC 1466   & 0.29 & 3.97 & 18.58 & 18.42 & 18.71 & 19.32 & 19.29 \nl
NGC 2257   & 0.13 & 3.78 & 18.39 & 18.55 & 18.52 & 19.10 & 18.99 \nl
Hodge 11   & 0.25 & 3.89 & 18.50 & 18.59 & 18.63 & 19.11 & .... \nl
\enddata
\end{deluxetable}

As discussed earlier, our $HST$ zeropoints have errors $\sim$
0.04$\,$mag, due to uncertain aperture and CTE corrections and the 
accuracy of the Holtzman \etal calibration.  To provide a comparison
with a ground-based reference, in Table 7 we include our HB magnitudes
and Walker's $<M_V(RR)>$ corrected by 0.04\mag for the offset between RR
Lyrae magnitudes and HB level (Gratton \etal 1997)\markcite{gr97}.
Added to this photometric uncertainty are our
uncertainties in the MS-fitting and in the differential reddening.
If we take 0.03\mag as an estimate of the uncertainty in the differential
reddening, then our error per cluster is $\sim$~0.07$\,$mag.  

We prefer the Pont \etal (1998)\markcite{pont98} M~92 distance as it
is based on a larger, better-selected sample of subdwarfs.  The distance
moduli for M92 and our clusters are listed in Table 7 as $\mu_{\odot,GC}$.
  Using this,
our LMC distance modulus from the average of our clusters' distances
is m$-$M$\sim 18.49$.  To illustrate the systematic uncertainty still
remaining because of the dispute over the local distance scale, we
include in Table 7 the distance moduli we derive using the Gratton
\etal (1997)\markcite{gr97} distance to M~92.  As for our measuring
error, the MS-fitting error and at least half of the zeropoint error
are random.  The systematic errors, such as the errors in the gain
ratios of the chips are mitigated because the center of the cluster is
not always on the same chip.  Our clusters are also not at the center
of the LMC.  We have attempted to correct for this by assuming that
the clusters lie in the HI disk (inclination $29^{\circ}$, PA of nodes
$-9^{\circ}$) as argued by Schommer \etal (1992)\markcite{sch92}.  
Our derived distances for
the LMC center are included in Table 7 ($\mu_{\odot,LMC}$).  This
changes our LMC distance modulus only slightly to 18.52$\,$mag.  In
fact, if the clusters lie in a disk, then our distance measurement is
relatively immune to changes in the assumed inclination and PA of the
nodes because NGC 1466 and NGC 2257 have about the same projected
distance from the LMC center but are on opposite sides of the LMC
center, while Hodge 11 is fairly close to the LMC center.  For our
final error budget, we include the 0.07\mag per cluster divided by
$\sqrt{3}$ plus 0.03\mag error in our averaged number due to uncertain 
geometrical corrections.  Also, Pont \etal quote 0.08\mag as their error in
the distance modulus to M92.  Our final (m$-M$)$_0$ is 18.46 $\pm$ 0.09$\,$mag.

\subsection{Blue Stragglers} 

All three of our clusters have a prominent blue-straggler star (BSS)
sequence.  To see how the frequency of BSS in these clusters compares
with that in GGCs, we used the specific frequency, $F_{BSS}$, defined by
Bolte, Hesser \& Stetson (1993)\markcite{b93} to be the ratio of the
number of BSS to the number of stars with $V < V_{HB}+2$.  Before we
can count stars in either category, we need to make two corrections.
First, even the 260s frames were long enough to saturate the brightest
RGB and AGB stars in the clusters and therefore they do not appear on
our CMDs. However, by looking at our images and star lists, it is easy
to find the number of bright stars ($\sim$ 15) that are not on the
list (Table 8).

\begin{deluxetable}{ccccccc} 
\tablenum{8} 
\tablewidth{0pt}
\tablecaption{Stellar Population Statistics} 
\tablehead{
\colhead{Cluster} & \colhead{\# ``BS''} & \colhead {\# ``BS''} & 
\colhead{\# RGB/HB/AGB} & \colhead {\# Saturated} & \colhead{\# RGB/HB/AGB}  
 & \colhead{Ratio}
\\ 
\colhead{} & \colhead{``cluster''} & \colhead {``field''} & \colhead{($V < V_{HB}+2$)} & \
colhead{Stars} & \colhead{($V < V_{HB}+2$)}  &
\colhead {$\rm Cluster Area$} \\
\colhead{} & \colhead{} & \colhead{} & \colhead{``cluster''} & \colhead{} & \colhead{``fie
ld''} & \colhead{$\overline {\rm Field Area}$}
}
\startdata
NGC 1466 & 73 & 0 & 721 & 17 & 26 & 1.41 \nl
NGC 2257 & 67 & 5 & 422 & 16 & 44 & 1.64 \nl 
\enddata
\end{deluxetable}
  The larger correction is due to field-star
contamination in the BSS region.  The analysis of Hodge 11 is severely
hampered by this, and we will not determine \fbss for this cluster.
For the ``BSS region'' in the CMD, we
adopted a box with magnitude boundaries at $V_{+0.05}-1.0$ and
$V_{+0.05}-3.0$ and color boundaries at \vmi$_{MSTO}-0.1$ and
\vmi$_{MSTO}-0.6$ (see Figures 15-16).
\begin{figure*}
\plotone{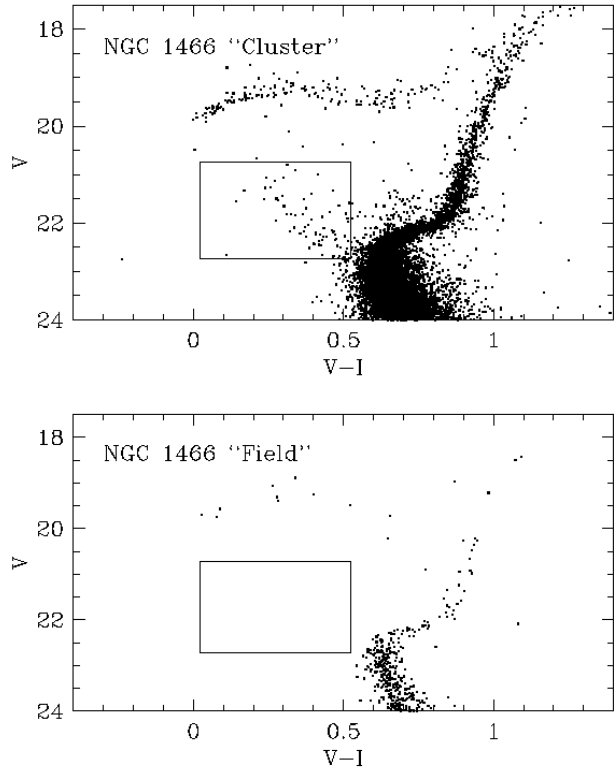} 
\caption{The upper part of the CMD of NGC~1466 with the
blue straggler box marked.  The upper panel shows the stars with r $<$
80\sec and the bottom with r $>$ 80$^{\prime\prime}$.  There is very little field
contamination evident in the ``field'' plot.}  
\end{figure*}
\begin{figure*} 
\plotone{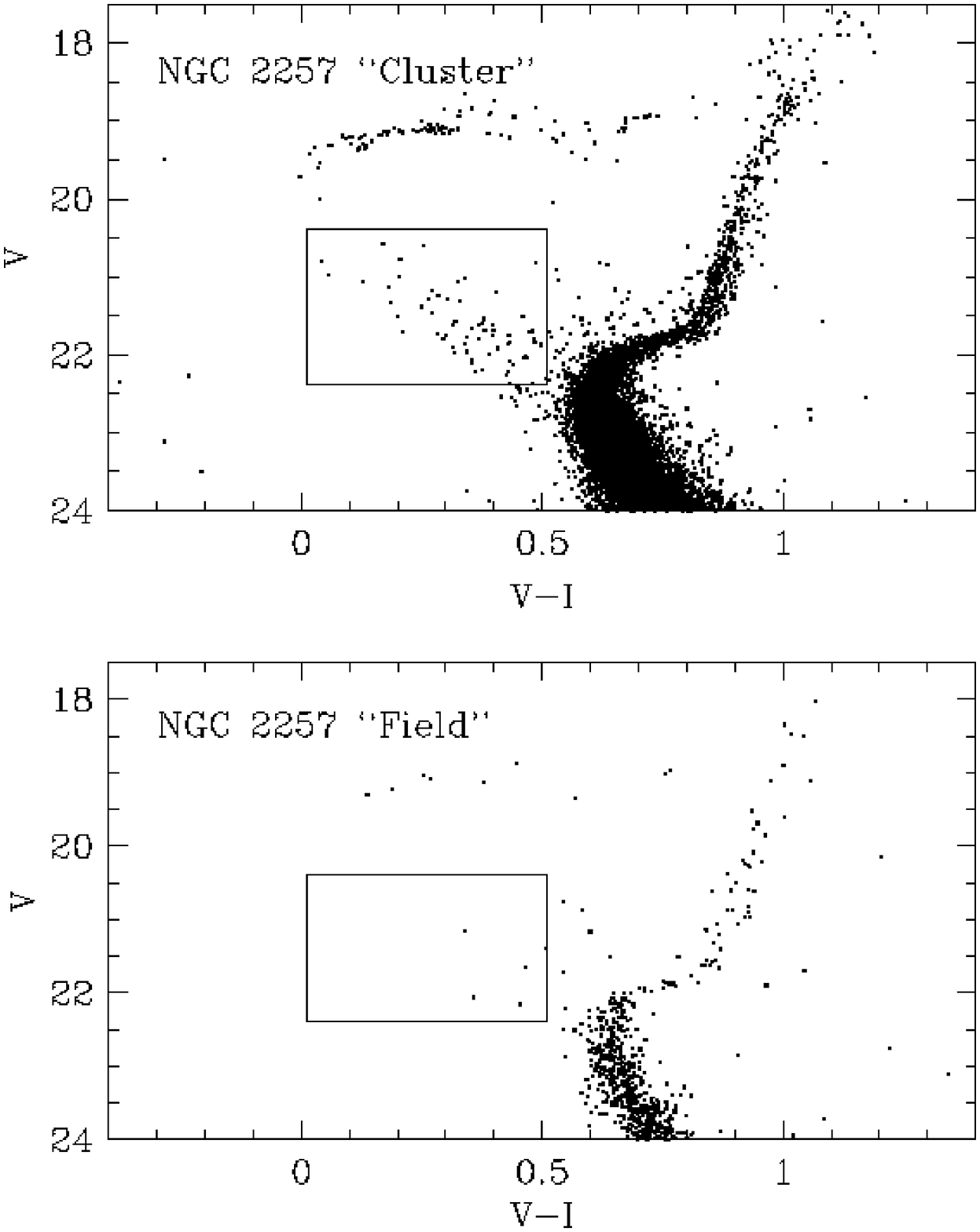}
\caption{The upper part of the CMD of NGC~2257 with the
blue straggler box marked.  The upper panel shows the stars with r $<$
80\sec and the bottom with r $>$ 80$^{\prime\prime}$.  While there are some stars
in the blue straggler box in the outer sample, the overall diagram shows
very little evidence for field contamination.  Therefore our derived
$F_{BSS}$ is most likely a lower limit.}  
\end{figure*} 
These boundaries are
somewhat arbitrary and are on the conservative side ({\it i.e.}, they
minimize contamination from the MSTO, but also may miss some BSS).
For NGC~1466 and NGC~2257, our total sample consists of the stars from
all the chips, with the shifts discussed in \S 4 applied (this step
makes very little difference).  To estimate the field contamination,
we divided our sample at r = $80^{\prime\prime}$.  The outer stars are not a true
field population, but will provide an upper limit on the number of
interlopers we might expect.  

In Figures 15 and 16, we show the upper
CMD of both the ``field'' and the ``cluster'' populations with the BSS
region marked.  The number of stars in each region is included in
Table 8.  (We note that all of the saturated stars belong to the ``cluster''
sample.)  NGC~2257 and NGC~1466 have \fbss of 0.14 and 0.10,
respectively.  For these numbers, we have only subtracted our
``field'' contamination from the blue straggler region.  We have not
considered contamination from field stars in the RGB+AGB+HB region.
Examination of the CMDs of stars further away from the cluster in
Walker (1992b) and Testa \etal (1995) show that the number of field
stars is about equal in both regions.  Since we have so many bright
RGB+AGB+HB stars compared to blue stragglers, possible contamination
in this region will not affect our results significantly.  In Table
9 we list the \fbss of some GGCs.  In this group we
include new numbers for Eridanus, Pal 3, Pal 4, and NGC 2419 based on the HST
observations of Stetson \etal (1999)\markcite{stet99} and Harris \etal
(1997)\markcite{har97} and using our definition of BSSs.
\begin{deluxetable}{llcll}
\tablenum{9}
\tablewidth{0pt}
\tablecaption{Blue Straggler Frequencies}
\tablehead{
\colhead{Cluster} & \colhead{$F_{BSS}$} & \colhead{Survey Region} & 
\colhead{c}  & \colhead {Source for $F_{BSS}$}
}
\startdata
\multicolumn{5}{c}{Milky Way Globulars} \nl
M 30 & 0.19 $\pm$ 0.04 & $r \leq 10 r_{core}$ & 2.40 & Yanny {\it et al.} 1994 \nl
NGC 6624 & 0.13 $\pm$ 0.14 & $r \leq 1.0 r_{core}$ & 2.15 & Sosin \& King 1995 \nl
47 Tuc & 0.07 $\pm$ 0.01 &$r \leq 3.0 r_{core}$ & 2.08 &  Guhathakurta {\it et al.} 1992 \nl
M 3 & 0.09 $\pm$ 0.02 & $r \leq 0.7 r_{core}$ & 1.89 & Guhathakurta {\it et al.} 1994 \nl
M 13 & $0.04-0.09$  & $r \leq 1.7 r_{core}$ & 1.44 & Cohen {\it et al.} 1997 \nl
NGC 2419 & 0.06 $\pm 0.005$ & $r \leq 8.2 r_{core}$ & 1.40 & Harris {\it et al.} 1997 \nl
Eridanus & 0.18 $\pm$ 0.05 & $r \leq 8.4 r_{core}$ & 1.14  & Stetson {\it et al.} 1999\nl
Pal 3 & 0.15 $\pm$ 0.05 & $r \leq 4.2 r_{core}$ & 0.96 & Stetson {\it et al.} 1999 \nl
Pal 4 & 0.12 $\pm$ 0.04 & $r \leq 4.9 r_{core}$ & 0.78 & Stetson {\it et al.} 1999 \nl 
NGC 5053 & 0.14 $\pm$ 0.03 & $r \leq 1.0 r_{core}$ & 0.77 & Nemec \& Cohen 1989 \nl

\multicolumn{5}{c}{LMC Globulars} \nl
NGC 1466 & 0.10 $\pm$ 0.01 & $r \leq 5.4 r_{core}$ & 1.42 & this paper \nl
NGC 2257 & 0.14 $\pm$ 0.02 & $r \leq 3.2 r_{core}$ & 1.10 & this paper \nl
\tablecomments{Structural parameters for GGCs from Trager {\it et al} (1993) and
for LMC from Mateo (1987).}

\enddata
\end{deluxetable}

\section{Discussion} 

\subsection{Blue Stragglers: \fbss} The globular clusters listed in
Table 9 have remarkably similar \fbss.  There are, of course,
systematic differences among the studies in their completeness, in
crowding effects and in the blue straggler definition.  However, while
there may be factors of up to two obscured by such problems, factors of ten
are very unlikely, based on comparing studies of the same cluster with
different completeness limits and crowding conditions.  We have chosen
mainly other HST studies to include in Table 9, which also makes the
sample more homogeneous.  The relatively narrow range in $F_{BSS}$,
compared to the $10^4$ range in cluster central densities, was noted
by Sosin \& King (1995)\markcite{sk95}.  They argued that as the
cluster density changes, different mechanisms ({\it e.g.}, stellar
collisions, tidal capture, merging of primordial binaries) become
important for the formation of blue stragglers.  Although the
efficiency of the various mechanisms varies from cluster to cluster,
the total production rate stays approximately constant.  Our new data
strengthen this point.  Eridanus has a \fbss at the high end
of the range and close to that of M~30.  M~30 is a post-core-collapse
cluster with a central mass density log($\rho_o$)= 5.26.  Eridanus,
on the other hand, is a sparse cluster at large Galactic
radius.  Its log($\rho_o$) is of order 0.5 (Webbink
1985)\markcite{web}.  The blue straggler specific frequencies for
NGC~1466 and NGC~2257 show that clusters belonging to other galaxies
have the same constraints on the rate of blue straggler formation as
those of the Galaxy.

\subsection{Ages and Age Distribution}
\subsubsection{Previous Results} The principle result of this study is
that these three LMC clusters are the same age as M~92 and M~3 to a
precision of 1.5 Gyr.  Mighell \etal (1996)\markcite{mig96} previously
reported that Hodge~11 and M~92 have the same age to within 10 to 21\%.
Testa \etal (1995)\markcite{t95}, based on a comparison of fiducial sequences and
$\Delta V$ values, found that NGC~2257 was about 2 Gyrs younger than
NGC 5897 and M~3.  In \S 5.4, we found that a modified $\Delta V$
method gives similar ages for M~3 and NGC~2257.  Part of the
discrepancy lies in the smaller $\Delta V$ value of 3.45 we find for
M~3 (Johnson \& Bolte 1998)\markcite{me98}.  We note that Stetson
\etal (1998) found a $\Delta V$ of 3.46 for M3.  Their sample included the
Johnson \& Bolte stars, but was about twice as large, including having
more HB stars.  

There are two other major studies of the ages of other old clusters in the
LMC.  Brocato \etal (1996)\markcite{bro96} argued on the basis of
$\Delta V$ measurements that NGC 1786, NGC 2210 and NGC 1841 have the
same age as the mean age of the GGCs, although their precision was
about $\pm$ 3 Gyrs.  Olsen \etal (1998)\markcite{ol98} (O98) found that the
ages of four old LMC clusters were the same as M~5 or M~55.  One
cluster, NGC 1835, was 2 Gyrs older than M~3.  Depending on the age
difference between M~3 and M~5, this could mean than NGC 1835 was coeval
with the other O98 clusters.  Another significant result from
their paper was the determination of the metallicities of their
clusters from the morphology of the RGB.  They found 
metallicities about 0.2 dex higher than those found by Olszewski
\etal (1991)\markcite{ols91}.

Before we can put all the above results together, including ours, we
need to consider possible age differences between the GGC samples used
to compare with the LMC ones.  Lee \etal (1994)\markcite{lee94} argued
on the basis of HB morphology that GGCs with $R_{GC} < 8$ kpc are on
average 2 Gyrs older than those between 8 and 40 kpc.  M~5 (6.4 kpc)
and M~55 (4.2 kpc) fall into the inner group while M~3 (8.5 kpc) and
M~92 (9.1 kpc) are members of the outer, possibly younger halo
(although note that the orbit of M~5 carries it into the outer halo).
Without resolving the issue of the chronology of GGC formation, we
will discuss results for the GGCs listed above.  Table
10 summarizes the recent relative ages for these four
clusters obtained from the literature.  
\begin{deluxetable}{llrrrl}
\tablenum{10}
\tablewidth{0pt}
\tablecaption{$\Delta$ Ages (Gyrs) for Comparison GGCs}
\tablehead{
\colhead{Source} & \colhead{M92} & \colhead{M55} & \colhead{M3} & \colhead{M5}
& \colhead{Method}}

\startdata
Richer {\it et al.} (1996) & 0.0 & 0.5 & $-1.7$ & .... & $\delta(V-I)$ \nl
Chaboyer {\it et al.} (1996b) & 0.0 &$-2.6$ & $-4.9$ & $-4.3$ & $\Delta V$ \nl
Gratton {\it et al.} (1997) & 0.0 & .... & .... & $-2.9$ & MS fitting \nl
Buonanno {\it et al.} (1998) & 0.0 & 1.1 & $-0.2$ & $-0.2$ & $\Delta V$, $\delta(V-I)$ \nl
\enddata
\end{deluxetable}
It is apparent that widely
varying opinions exist on the age spread in the Galactic halo.  Part
of this is due to observational uncertainties.  Our value for the HB
magnitude of M~92 (15.18) is considerably fainter than the 14.96 used
by Chaboyer \etal (1996b)\markcite{chab96b} and Gratton \etal
(1997)\markcite{gr97}.  If our value is correct, then the age gap
between M~92 and the other clusters will be reduced.  However,
although uncertainties still exist, the outer halo clusters M~92 and
M~3 are apparently not younger than the inner halo M~5 and M~55.  We 
can directly compare our LMC fiducials to the inner-halo GGC
M~55, which has a metallicity, [Fe/H]$=-1.85$ (ZW84), about the same
as our LMC clusters.  G. Mandushev kindly sent us the $V$, \vmi
photometry of Mandushev \etal (1996)\markcite{mand96}.  We choose the
NGC~2257 fiducial as representative of our LMC clusters.  Figure
17 shows that the two cluster fiducials match very
closely throughout the CMD. As is the case for the LMC$-$M~92
comparison, there is at most a small, $<1.5$ Gyr, age difference
between NGC~2257 and M~55.
\begin{figure*}
\plotone{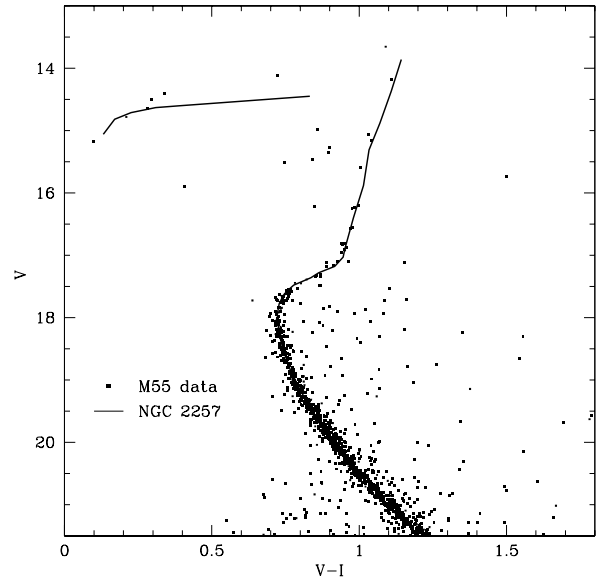}
\caption{NGC~2257's fiducial compared to the M~55 data of Mandushev
\etal (1996).  The overall agreement between these two clusters of
comparable metallicity shows that there is not a large age difference
between them.}  
\end{figure*}

\subsubsection{Implications for the Formation of the Local Group}

The LMC has been suggested as the prototype of a source for
present-day GGC.  One of the scenarios advocated by Searle \& Zinn
(1978)\markcite{sz} was that clusters formed in ``fragments'', perhaps small
gas-rich irregulars, that evolved and interacted over the span of
several billion years.  In
the particular case of the LMC, Zinn (1980b, 1993) has suggested that
satellites like the LMC contributed their GCs to the MW halo. Previous
observations of the LMC clusters revealed that they are similar to the
GGC in many respects (see Suntzeff \etal 1992\markcite{s92}).
Suntzeff \etal pointed out that the GGC system could have been made by
merging $\sim$ 8 LMC-like objects.  Van den Bergh (1996)
found, on the other hand, that the metallicities of the old LMC
clusters in the Suntzeff \etal sample were, on average, lower than
those for clusters in the MW's outer halo.  The increased metallicity
estimates by O98 for five of the old LMC clusters would make this
discrepancy smaller.  Van den Bergh \& Morbey (1984)\markcite{vdb94}
and others also argued that since the ellipticity of the LMC clusters
is greater than in the MW clusters, it was unlikely that the MW had
captured LMC clusters. The LMC clusters have a larger average
ellipticity regardless of the age range considered, although at the
time, accurate ages were not available.  Here we consider only those
LMC clusters that have been confirmed as old and that have ellipticity
measurements by Frenk \& Fall (1982)\markcite{ff82} and those MW
clusters that are not heavily reddened (E(\bmv)$< 0.32$) (Harris
1992).  A two-sided K-S test can reject only at the 77\% level the
hyphothesis that these two groups were drawn from the same
distribution.  Goodwin (1997)\markcite{good97} notes that cluster
ellipticity can depend more strongly on the tidal field of the parent
galaxy than it does on intrinsic cluster properties. For the specific
case of the three LMC clusters considered here, each of them has a
small eccentricity, consistent with the distribution seen in the GGCs.

Another similarity between the old clusters of the Galaxy and LMC is in the
 distribution
of HB morphology with [Fe/H].  In the outer halo of the Galaxy, the HB
morphology of GGCs is on average too red for the cluster metallicity
(the ``young halo'' of Lee \etal 1994). Zinn (1993) showed that 
the best match to the outer halo GGC properties in terms of [Fe/H]
and HB morphology were the clusters of the LMC and SMC.  (In fact, the
SMC globular NGC 121 was the first ``second parameter'' globular cluster
noticed (van den Bergh 1967)).  He
suggested that the majority of the inner halo clusters formed in the
overall collapse of the Galaxy, but that the outer halo was mostly
accreted from relatively large satellites.

The new $HST$ results allow us to make a few comments about the
scenario in which LMC-like galaxies are absorbed by the Galaxy and
contribute their stars to the Galactic halo. On the basis of age, the
{\it old} LMC clusters would be indistinguishable from existing GGCs,
at least our comparison clusters in the intermediate halo.  With the
recent age estimates and, for the O98 clusters, revisions in [Fe/H]
estimates, it is also possible to improve on the [Fe/H] -- HB
morphology comparison between the LMC clusters and different GGC
populations.  In Figure 18, we plot our version of HB-type versus
[Fe/H] originally shown in Lee (1990)\markcite{lee90}.  
\begin{figure*}
\plotone{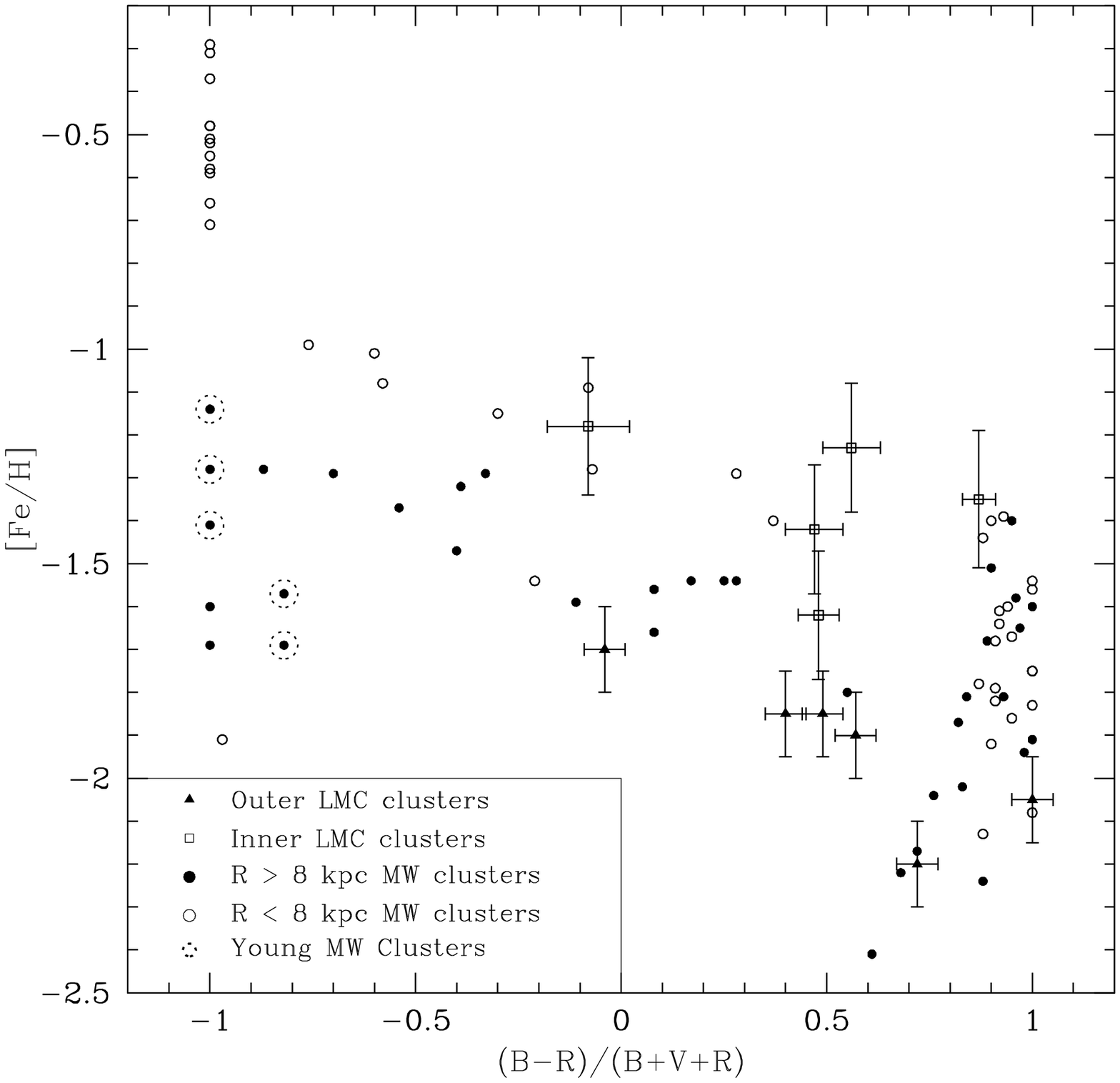}
\caption{HB-type versus [Fe/H] for LMC and Galactic globular clusters.
The HB-type is defined to be the number of blue stars minus the number
of red stars divided by the total number of stars on the HB, including
variables.
The Galactic clusters have been divided into two groups, as suggested
by Lee \etal (1994).  Those MW clusters with R $<$ 8 kpc tend to have
bluer HBs than similar metallicity ones with R $>$ 8 kpc.  The LMC
clusters display the same trend when the inner and outer clusters are
compared.  The data for inner LMC clusters are from Olsen \etal (1998), 
while the data for the outer ones are from Walker (1992b).  The data
for the MW come from the complilation in Lee \etal (1994).
We have marked with a dotted circle those MW clusters that are most likely 
truly young clusters, based on their turnoff magnitudes.  They are 
concentrated at even redder HBs than the LMC clusters.}  
\end{figure*} 
The O98
clusters at their revised metallicities show nice agreement with the
inner halo clusters of the Galaxy, while our clusters fall among the
outer halo clusters, as originally noted by Zinn (1993).  Since the
O98 clusters were not well-studied before HST and had no HB-type, they
were not included in the analysis of Zinn (1993).  However, with the
new data, we can see that if the LMC contributes all of its GCs to the
MW halo, it will contribute clusters like those in the inner halo as
well, in about equal proportions.

There is another interesting, though tentative point.  If we accept
the O98 metallicities, then, as is the case for the GGC,
the LMC clusters also separate into two
groups in the HB-type versus [Fe/H] plot.   One group,
composed of the O98 clusters, has higher [Fe/H]'s but not redder HBs 
than the other,
which includes all the clusters studied by Walker (1992b), including
NGC~1466, NGC~2257 and Hodge~11.  The LMC groups also mimic the GGC in their
spatial distribution differences.  The O98 clusters are concentrated
toward the center of the LMC.  They all have projected angular
separations $<$ 2.5\Deg from the LMC center, whereas the Walker sample
has no clusters closer than 2.5\Deg from the center.
  Based on our analysis in \S 6.2.1, where we discussed the age range
among all the LMC clusters studied with HST, it is possible that age
in this case in not the second parameter, a point also made by Da
Costa (1999)\markcite{da99}.  A picture where the LMC builds up its
outer GCs by accreting satellites is also less appealing, because of
the mass of the LMC is not much larger than the the mass of the
smallest satellites known to have their own globular cluster systems.
An additional clue to the formation of the LMC cluster system comes
from the possible radial abundance gradient, if the O98 metallicities
are correct (Da Costa 1999).

\begin{figure*}
\centerline{
\psfig{figure=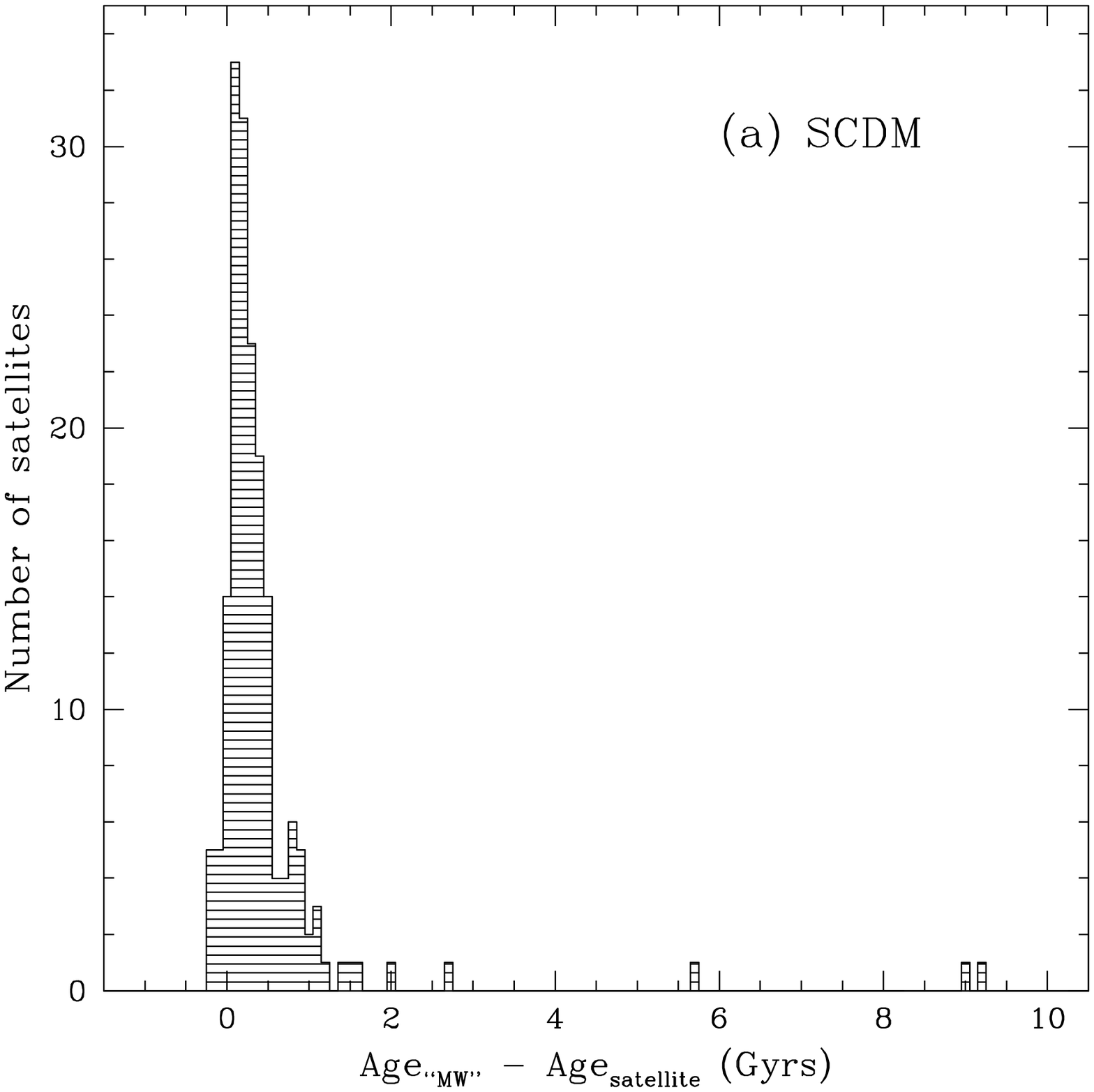,width=2.5truein,angle=0}
\psfig{figure=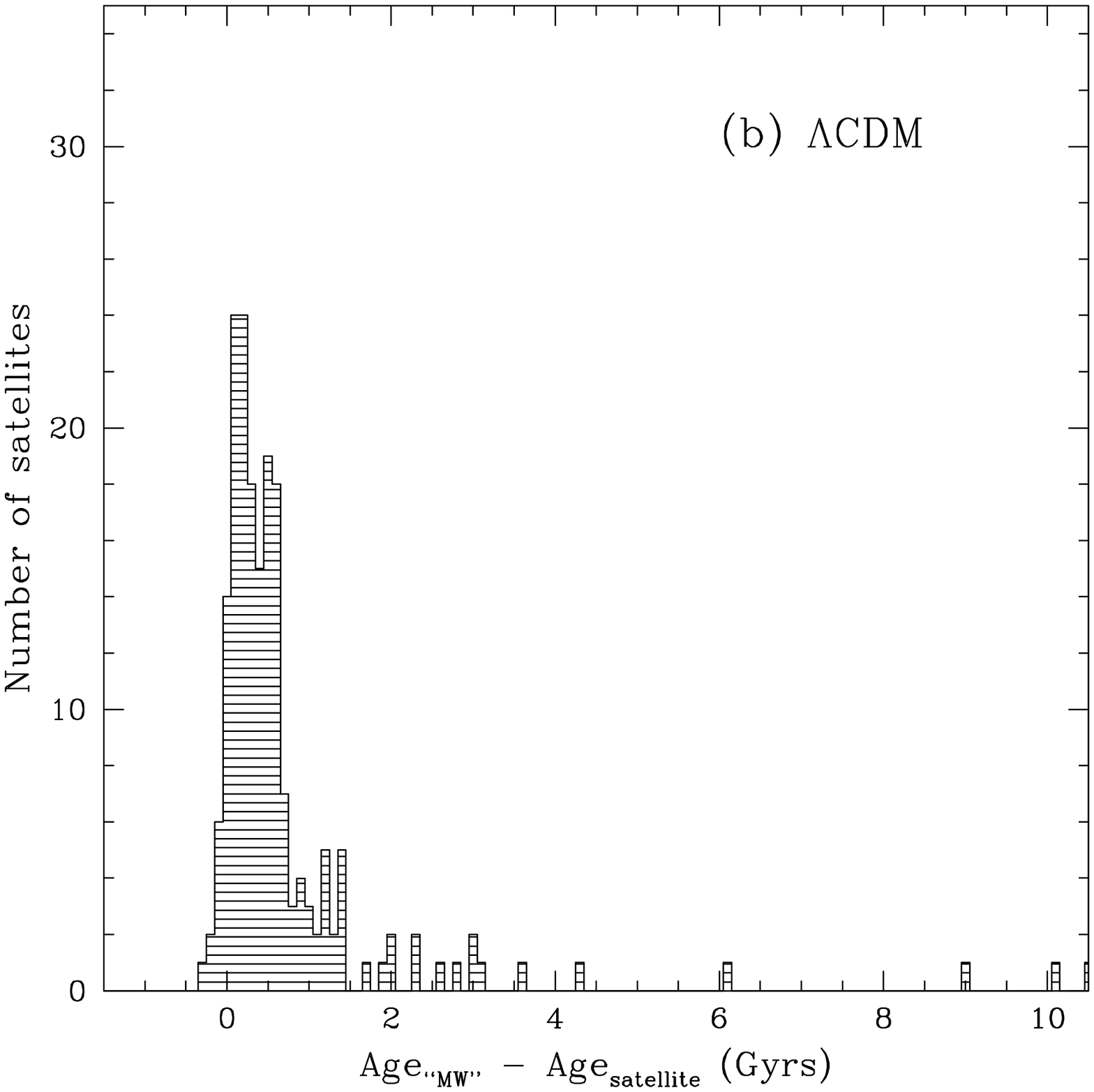,width=2.5truein,angle=0}
}
\caption{A histogram of the time lapse between the collapse of the first
progenitor of the parent galaxy and of each of its bright ($M_B-5{\rm logh} 
\leq -15$) satellites for our ``Milky Way'' halos in (a) a standard CDM universe 
and (b) a $\Lambda$CDM universe.  At the level that we can determine
age differences, these two distributions are essentially the same.}
\end{figure*}

\subsubsection{The Big Picture} Our most fundamental result is that
the oldest LMC clusters are the same age as the old GGC.  What does
this mean for our understanding of galaxy formation?  On one extreme,
if we imagine the Galaxy being made completely via the accumulation of
LMC-sized fragments (and if we accept the argument of Stetson \etal
(1996)\markcite{stet96} that most GGC are essentially coeval), then this
coincidence in age would be telling us that the collapse epoch of all
the original fragments was the same to 1.5 Gyr or so.  In the opposite
extreme, in which only a small fraction of the current-day Galaxy was
acquired through accumulation of small companions, our result is then
telling us that the formation epoch for the Milky Way out to the
intermediate halo is the same to $\sim 1.5$ Gyr as for the LMC, a
galaxy ten times less massive.

We discuss briefly here this result in the context of one (popular)
galaxy formation scenario -- hierarchical clustering ({\it e.g.}, Blumenthal
\etal 1984)\markcite{bl84}. It is not yet possible to fully simulate the 
processes that lead to galaxy formation. In particular, the question of when globular
clusters form within hierarchical clustering models is not yet accessible to 
simulations. Nevertheless, with the reasonable assumptions
that globular clusters were formed within larger structures
(e.g. a proto-LMC) and were among the first objects formed
after the gas began to collapse, we can investigate
whether the current hierarchical formation models can
be consistent with our ``fossil record'' data. We note that as the models
become more complete and sophisticated, this discussion may become rapidly
dated. Nevertheless, as the data for globular cluster ages continue to improve,
they will become increasingly important in testing models such as these.

As an example of what hierarchical clustering models predict for the
formation epoch of LMC-like galaxies, we examined the properties of
galaxies in 100 ``Milky Way'' halos calculated by the semi-analytic
method of Somerville \& Kolatt (1999) and Someville \& Primack (1999) 
\markcite{som98a}\markcite{som98b}.  These semi-analytic models
are a powerful way to study galaxy formation by combining merging
histories for dark-matter halos based on extended Press-Schechter
theory (Bower 1991\markcite{bow91}; Bond \etal 1991\markcite{bond})
with simple prescriptions for such physical processes as gas cooling, star
formation and supernova feedback.  The advantage of using this method
is that the formation and evolution of many systems in many different
cosmologies can be calculated in a reasonable period of time.

  Placing GC formation in this scenario is difficult,
because there is no agreed-on model of globular cluster formation and
these models are not hydrodynamical calculations that can probe
globular cluster formation.  However, since the globular clusters are
slightly metal-enriched and since models of globular cluster formation
generally start with cold molecular gas ({\it e.g.}, Harris and Pudritz
1994; Elmegreen \& Efremov 1997)\markcite{har94}\markcite{e97}, we can
place the epoch of formation just after the gas can cool and start to
form stars.
  When this happens depends on the mass of the dark matter
halo.  Because of the meta-galactic UV field, gas is
prevented from cooling until its dark matter halo has reached the
critical mass when its gas is dense enough to be self-shielded from
ionizing photons.  Simulations have shown that while this critical mass
depends on redshift, it is not sensitive to such details as the shape
of the UV-spectrum and the inclusion of star formation, especially at
the level of precision we require ({\it e.g.}, Weinberg, Hernquist \&
Katz 1997)\markcite{whk}.

For each present-day galaxy in our ``Milky Way'' halo catalog, we have the
redshift ($z_{coll}$) when any of the halos in that particular galaxy's merging
tree reached that critical mass in dark matter (the ``first
progenitor'').  We assume that this marks the initial stage of globular
cluster formation.  We confine our results here to those satellites with
$M_B < 5{\rm log(h)} - 15$\footnote[2]{h=$H_0/100$}, which includes the 
LMC and SMC, but not the dwarf spheroidals.


To
illustrate the range of separation in time of this epoch, we calculated 
the difference in $z_{coll}$ between the
Milky-Way type galaxy and each of its satellites in the 100 model
halos.  The results are plotted in Figures 19a-b for
two different cosmologies, standard cold dark matter (SCDM) with
$\Omega_{matter}$=1.0 and $H_0$=50 km s$^{-1}$ Mpc$^{-1}$ and cold
dark matter with a cosmological constant ($\Lambda$CDM) with 
$\Omega_{\Lambda}=0.7$, $\Omega_{matter}=0.3$, and $H_0$=70 km
s$^{-1}$ Mpc$^{-1}$.
While there is a tail to large age differences, there is a strong peak
at small age differences in both cosmologies.  We cannot measure age
differences much smaller than 1 Gyr.  Therefore, if we assume that the
initial epoch of globular cluster formation in a galaxy happened at
about the same time as the collapse of the first progenitor of that
galaxy, then our result that the oldest LMC and the Galactic clusters
formed at the same time $\pm$ 1.5 Gyr is completely consistent with
current hierarchical clustering models.  As Harris \etal
(1997)\markcite{har97} pointed out, this apparent synchronicity between
the LMC and Galaxy, and within the outer halo of the Galaxy, is due to
a number of {\it individual} dark matter halos approximately
simultaneously reaching a critical mass so that star formation can
start.  The SMC provides a counter-example as it apparently began its
cluster formation $\sim$ 2 Gyr later ({\it e.g.}, Stryker, Da Costa, \&
Mould 1985\markcite{st85}; Mighell, Sarajedini, \& French
1998\markcite{mig98}).  In our hierarchial clustering models, it would
be in the tail of the ``Milky Way''-satellite age distribution.

We conclude that hierarchial clustering models, at least as described
by current semi-analytical codes, are consistent with both constraints
mentioned at the beginning of this section, though it is
slightly harder to satisfy the first constraint for a
$\Lambda$CDM case.  While many LMC-size galaxies
form at about the same time, there is dispersion in $z_{coll}$,
making it less likely to have all the original fragments collapse at the
same time.  On the other hand, the mass difference between
the Milky Way and the LMC is not enough for it to be unusual for the
collapse of the first progenitor to happen at the same time for both
galaxies.

\acknowledgments We would like to thank Dan Kelson, Raja
Guhathakurta, Ata Sarajedini, and Eric Sandquist for useful
suggestions and computer programs.  Our thanks to V. Testa, A. Walker,
K. Mighell, and G. Mandushev for providing their data on these
clusters.  J. A. J.  acknowledges partial support from an NSF Graduate
Student Fellowship.  M. B. is happy to acknowlege support for this
program from NASA grant GO-5897.01 administered through STScI and
NSF grant AST 94-20204.

\clearpage
\end{document}